\documentclass{article}
\usepackage{amsmath}
\usepackage{mathrsfs}
\usepackage{geometry}
\usepackage{cancel}
\usepackage{multicol}
\usepackage{tikz}
\usepackage{breqn}
\usepackage{braket}
\usepackage{soul}
\usepackage{tikz-feynman}
\usepackage{graphicx}
\usepackage[english]{babel}
\usepackage{subcaption}
\usepackage{float}
\usepackage{lipsum}
\usepackage{hyperref}
\usepackage{float}
\usepackage{multirow}
\hypersetup{
    colorlinks=true,
    citecolor=cyan,
    linkcolor=blue,
    urlcolor=blue}
\usepackage{amssymb}% http://ctan.org/pkg/amssymb
\usepackage{pifont}% http://ctan.org/pkg/pifont
\usepackage{authblk}
\def\beq{\begin{equation}}
\def\eeq{\end{equation}}
\def\barr{\begin{array}}
\def\earr{\end{array}}
\def\dis{\displaystyle}

\newcommand{\be}{\begin{equation}}
\newcommand{\ee}{\end{equation}}
\newcommand{\bea}{\begin{eqnarray}}
\newcommand{\eea}{\end{eqnarray}}

\newcommand{\bi}{\begin{itemize}}
	\newcommand{\ei}{\end{itemize}}

\newcommand{\Slash}[1]{{\ooalign{\hfil/\hfil\crcr$#1$}}}
\title{Axi-Higgs portal Dark Matter via Wess-Zumino mechanism}
\author[]{Akshay Anilkumar}
\author[]{Mathew Thomas Arun}
\author[]{Arjun S. Nair}
\affil[]{School of Physics, Indian Institute of Science Education and Research, Thiruvananthapuram-695551, Kerala, India}
\date{}                     
\begin{document}
  \maketitle

  \begin{abstract}
    
    We study the axion portal between the visible and the dark sector, where the dark matter is charged under an abelian extension of the Standard Model. In general, such models are anomalous and are rendered gauge invariant by a St{\"u}ckelberg axion through Wess-Zumino/Green-Schwarz mechanism. Scenarios such as this naturally exist in TeV scale string theory completions of Standard Model. This axion mixes with other Goldstone bosons in the model to give a physical axi-Higgs which becomes massive upon breaking the anomalous gauge group. Such axi-Higgs fields charged under the anomalous symmetry act as mediators for the dark matter annihilation to Standard Model particles and can lead to an efficient freeze-out mechanism. Here, we show that the St{\"u}ckelberg axion, and the resultant axi-Higgs, with its appropriate shift symmetry cancels the quantum anomalies and also generates the observed relic density for the dark matter. Moreover, we show that the relevant parameter space in our model, where photon production dominates, is safe from {\it Fermi}LAT, Cherenkov Telescope Array, and H.E.S.S. indirect detection experiments.
  \end{abstract}
\section{Introduction} \label{sec:intro}
The origin of dark matter is probably one of the most relevant discussions in High Energy Physics and cosmology, currently. Though it is overwhelmingly present in astrophysical and cosmological contexts~\cite{Bertone:2004pz,WMAP:2008lyn,Planck:2015fie, XENON:2018voc}, its lack of evidence in terrestrial experiments poses a conundrum. This, along with the absence of a natural dark matter candidate in the Standard Model put forth compelling reasons for going beyond the current understanding of particle physics. One of the most successful paradigms in DM modelling is the thermal freeze-out scenario~\cite{Vysotsky:1977pe,Sato:1977ye,Lee:1977ua, Hut:1977zn,Steigman:1979kw,Kolb:1985nn,Scherrer:1985zt,Griest:1990kh,Jungman:1995df,Arcadi:2017kky}. Here, the dark matter density decouples from thermal equilibrium with visible sector when its annihilation rate becomes of the order of Hubble expansion rate, at some temperature during the evolution of our universe. The annihilation of the DM particle to the Standard Model is facilitated via an interaction between the dark and visible sectors, which is generated as a result of modifying either the gauge group or the space-time symmetries in the Standard Model. For such dark matter candidates, constraints from various direct and indirect detection experiments and their production at the {\it Large Hadron Collider} (LHC) become relevant. Most such extensions of SM with spontaneous breaking of the symmetry contain light pseudo-Goldstone bosons which could be searched for at colliders and can themselves mediate an efficient portal between these dark and visible sectors~\cite{Nomura:2008ru,Freytsis:2010ne,Essig:2013lka,Dolan:2014ska,Antel:2023hkf, Fitzpatrick:2023xks, Armando:2023zwz,Ghosh:2023tyz}.

Embedding Standard Model within TeV scale string theory has been of immense interest in the context of {\it Naturalness}~\cite{Arkani-Hamed:1998jmv,Antoniadis:1998ig}, which implies that the fundamental scale of gravity must be much smaller than the Planck mass. Such models contain a stack of N identical parallel D-branes that generate the $U(N)$ non-abelian gauge symmetry, while, chiral matter is understood as strings stretching between two stacks of intersecting D-branes. These string models are interesting since they can be used to construct UV completions of SM with the low energy effective field theory predicting signatures like Regge resonances in particle colliders. Various such intersecting D-brane models have been proposed~\cite{Ibanez:1998qp,Blumenhagen:2006ci, Blumenhagen:2005mu, Antoniadis:2001sw, Anchordoqui:2022kuw,Anastasopoulos:2022wob,Armillis:2007tb,Anastasopoulos:2022ywj} which contain extra anomalous $U(1)$s with baryon and lepton number. In our paper, we focus on the possibility of fermionic dark matter candidate in a leptophilic anomalous $U(1)$ model. Such a leptophilic model can be realised with 5 intersecting stacks of D-branes~\cite{Armillis:2007tb,Antoniadis:2021mqz,Anastasopoulos:2022ywj}. These models are interesting as they contain anomalous triple gauge boson interaction that contribute to $g-2$ of muon and predict lepton flavour non-universality. In this work, we consider the Wess-Zumino mechanism, which is required to make the model anomaly free, and ask whether this term can provide an efficient portal for the fermionic dark matter candidates to annihilate.

Though the gauge sector of Standard Model is enlarged with multiple abelian gauge groups in intersecting D-branes models, they are plagued with gauge anomalies. While in beyond Standard Model field theoretic scenarios, discussed widely in literature, the most common method to cancel such tri-linear gauge anomalies is through suitably distributing the charges among fermions of each generation, several proposals from string theory~\cite{Ibanez:1998qp} and extra dimensions have brought in new perspectives regarding the cancellation of these anomalies with higher dimensional operators. These proposals include the introduction of a Wess-Zumino and a Chern-Simons term in the Lagrangian, leading to a local Wess-Zumino (WZ) mechanism~\cite{Wess:1971yu,Anastasopoulos:2008jt} in field theory and non-local Green-Schwarz (GS) mechanism~\cite{Green:1984sg,Green:1996dd,Anastasopoulos:2006cz} in string theory. In such cases, the reducible anomalies vanish through the introduction of new counterterms. Though WZ and GS mechanisms are closely related~\cite{Coriano:2008pg,Armillis:2008bg}, in the former the anomaly cancellation occurs at the Lagrangian level by the shift symmetry of a St{\"u}ckelberg axion that couples to the Wess-Zumino term, while the latter mechanism cancels the longitudinal components of the anomalous vertex, thus restoring the gauge invariance. Since in the GS mechanism, the St{\"u}ckelberg axion is integrated out to generate non-local forms, for the purpose of this work, we consider the Wess-Zumino mechanism for anomaly cancellations and generation of interactions.

In this article, we study the dark portal in low-energy intersecting D-brane model like the one constructed in~\cite{Antoniadis:2021mqz,Anastasopoulos:2022ywj} where the Standard Model gauge group gets extended with a local anomalous $U(1)^{\prime}$. In such models, understanding the importance of the Wess-Zumino mechanism is necessary at the level of model building. Since irreducible anomalies are absent in this anomalous $U(1)^{\prime}$ extension, the gauge invariance can be restored by the  WZ anomaly cancellation mechanism by including a St{\"u}ckelberg axion that couples to 4-form $F_I\wedge F_J$ of the gauge fields ($I,J$). In this set-up, we consider 3 stacks of intersecting  D-branes, giving rise to $U(3) \times U(2) \times U(1)$ gauge symmetry at the low energy limit, and a $U(1)^{\prime}$ brane in the bulk so that this $U(1)^{\prime}$ charge does not contribute to the hypercharge in Standard Model. The Standard Model leptons doublets and singlets are realised to be open strings between the $U(2)$ and $U(1)$ D-brane to the $U(1)^{\prime}$ brane respectively. Since we are interested in leptophilic models, and to keep the discussion simple, we will not address the arrangement of the quarks here. 
We also extend the Standard Model matter sector to include a fermionic dark matter candidate emerging from an open string between the $U(1)^{\prime}$ brane and a dark brane the bulk. Apart from the regular Higgs scalar, we also consider a complex scalar charged under the $U(1)^{\prime}$. Though it might seem that the St{\"u}ckelberg axion in the WZ mechanism is {\it eaten} away by the new $Z^{\prime}$ boson, in the presence of the complex scalar field, the axion mixes with other Goldstone bosons to produce a physical Goldstone field~\cite{Coriano:2007fw}.

We will, here, show that this physical Goldstone boson (namely axi-Higgs) gets massive and mediates an efficient portal between the dark and the visible sectors to generate the observed dark matter relic density. In fact, the most dominant channels include $DM+DM \to W^+W^-, \gamma Z, \gamma \gamma$ and $ZZ$, for $\sim \mathcal{O}(10\text{GeV})$ dark matter masses, generated by the Wess-Zumino term.

These axion-gauge boson interactions are constrained by several astrophysical observations and terrestrial experiments. The photon coupling of axions in meV and keV scales are strongly constrained by CAST~\cite{CAST:2017uph} helioscope and the ``energy-loss argument"~\cite{Caputo:2024oqc}. Similarly, for axions in the 100 MeV mass scale it is constrained by the bounds from supernova cooling of SN1987a~\cite{Payez:2014xsa,Jaeckel:2017tud} and beam dump experiments~\cite{Blumlein:2013cua,Bjorken:1988as,Riordan:1987aw}. Moreover, for axions in the MeV mass range, the axion-W boson coupling ($g_{bWW}$) receives additional constraints from rare decay searches, particularly from the E787 and E949 experiments~\cite{BNL-E949:2009dza,Izaguirre:2016dfi,Alonso_lvarez_2019} and CHARM experiment~\cite{CHARM:1985anb}. Since the emphasis in this work is on the axi-Higgs portal to electro-weak gauge bosons, the axi-Higgs we consider are heavy ($m_{G_{1}} > \mathcal{O}(10\text{ }{\rm GeV}))$. While the most stringent bounds are for the couplings of lighter axions, for axions in the mass range of our interest, LEP constraints from the $Z \to \gamma\gamma$ and $Z \to \gamma\gamma\gamma$ searches put a limit of $g_{b\gamma\gamma} \lesssim 7\times 10^{-3} \text{ GeV$^{-1}$ for } m_{G_{1}} \text{ of } \mathcal{O}(1\text{ }{\rm GeV})$, $g_{b\gamma\gamma} \lesssim 9\times 10^{-4} \text{ GeV$^{-1}$ for } m_{G_{1}} \text{ of } \mathcal{O} (10 \text{ }{\rm GeV})$, and $g_{b\gamma\gamma} \lesssim 10^{-3} \text{ GeV$^{-1}$ for } m_{G_{1}} \text{ of } \mathcal{O}(100\text{ }{\rm GeV})$ ~\cite{Jaeckel:2015jla,Mimasu:2014nea}. Similarly, axions in the mass range 1 GeV to 1000 GeV also face constraints from LHCb~\cite{Benson:2314368} and BaBar~\cite{BaBar:2011kau}. These experiments together set a bound of $g_{b\gamma\gamma} \lesssim 2 \times 10^{-4}$ GeV$^{-1}$~\cite{Alonso_lvarez_2019}. The $Z \to \gamma + hadron$ searches at L3 collaboration~\cite{L3:1992kcg} of LEP constrains $g_{bZ\gamma} \lesssim 2 \times 10^{-4}$ GeV$^{-1}$  for $10\text{ }{\rm GeV}\lesssim m_{G_{1}} \lesssim 100\text{ }{\rm GeV}$. For even higher $m_{G_{1}}$ values, in the sub-TeV and TeV scale, the major constraints on the axion-gauge boson couplings arise from LHC~\cite{Jaeckel:2012yz,Mariotti:2017vtv}, where it is searched in the channel $\sigma(pp \to axion)\times BR(axion \to \gamma \gamma)$, limiting the $g_{b}\gamma\gamma$ coupling to $< 10^{-4}\text{ }{\rm GeV^{-1}}$. 

To explain the framework, we consider two models, anomalous $U(1)^{\prime}_{e_\mu L_\mu -e_\tau L_\tau}$ and anomalous $U(1)^{\prime}_{\gamma_5(L_e+L_\mu+L_\tau)}$. As mentioned before, these models could arise in intersecting D-brane models, where, the leptons arise from open strings extending from $U(2)$ and $U(1)$ branes to the $U(1)^{\prime}$ brane in the bulk. Here, we do not include the interactions of quarks for simplicity. In these scenarios, we show that the models satisfy the observed relic density bound and the interested parameter space is well within the indirect detection results of {\it Fermi}LAT~\cite{Atwood_2009, Ackermann_2015}, Cherenkov Telescope Array~\cite{CTAConsortium:2010umy,CherenkovTelescopeArray:2024osy} and H.E.S.S~\cite{Abdallah_2018}.

This article is organised as follows. In Sec.\ref{sec:WZ} we briefly recapitulate the Wess-Zumino mechanism for gauge anomaly cancellation and the mixing of St{\"u}ckelberg axion with the Goldstone boson. In Sec.\ref{sec:model} we describe a generic anomalous $U(1)^{\prime}$ extension to the Standard Model, along with a complex scalar field and a fermionic dark matter. We then consider the anomalous $U(1)^{\prime}_{e_\mu L_\mu -e_\tau L_\tau}$ and the anomalous $U(1)^{\prime}_{\gamma_5(L_e+L_\mu+L_\tau)}$ and derive the axi-Higgs couplings with gauge bosons and fermions. In Sec.\ref{sec:relic} we compute the relic density of dark matter in these models and in Sec.\ref{sec:indirect} we constraint our model with data from {\it Fermi}LAT, Cherenkov Telescope Array and H.E.S.S.. Finally, we summarise our results in Sec.\ref{sec:conclusion}.

\section{Wess-Zumino anomaly cancellation mechanism and axi-Higgs}
\label{sec:WZ}
Below, we will briefly discuss the Wess-Zumino anomaly cancellation mechanism~\cite{Wess:1971yu}.

Let's consider an $U(1)$ gauge field $Z^{\prime}_\mu$ and a fermion field that transform under this gauge group. Such models can lead to quantum anomalies if the fermion interaction with gauge boson is chiral in nature. A remedy to this issue is given by the St{\"u}ckelberg axion with the symmetry $b \to b+ f_b \alpha(x)$ under the gauge transformation. This action is given by, 
\begin{eqnarray}
  S= \int d^4x  \mathcal{L} =  \int d^4x \{ -\frac{1}{4} Z^{\prime}_{ \mu \nu} Z^{\prime \mu \nu} +\frac{1}{2}(\partial_\mu b + f_b Z^{\prime}_\mu)^2  + \bar{\psi} i \gamma^\mu (\partial_\mu +i g \gamma^5 Z^{\prime}_\mu) \psi + \frac{g^3C^{BBB}}{8 \pi^2 f_b}  \ b \ Z^{\prime}_{\mu\nu} \Tilde{Z}^{\prime \mu \nu} + L_{R_\xi} \} \ ,
 \label{eq:WZLagrangian}
\end{eqnarray}
where $C^{BBB}$ is an effective coupling of the St{\"u}ckelberg axion with the Wess-Zumino term and $L_{R_\xi}$ is the generalised $R_{\xi}$ gauge fixing term given by $L_{R_{\xi}}= -\frac{1}{2 \xi}(\partial_\mu Z^{\prime \mu} -\xi f_b b)^2$. This term cancels the mixing, $Z^{\prime \mu} \partial_\mu b$, arising from the kinetic part of the St{\"u}ckelberg axion. To see how this Wess-Zumino interaction term keeps the current conserved, lets consider the partition function,
\begin{equation}
Z = \mathcal{N} \int d\bar{\psi}d\psi e^{-S} \ .
\end{equation}
Using the Fujikawa method~\cite{Fujikawa:1979ay,Fujikawa:1980eg} of deriving the anomaly from the path integral, upon gauge transformation, it can be seen that the anomalous part arises from the Jacobian of the measure $d\bar{\psi}d\psi$ in the path integral. Thus, for the massless fermion field $\psi$, along with $b \to b+ f_b \alpha(x)$, the Ward identity becomes,
\begin{equation}
0=\frac{\delta Z}{\delta \alpha(x)}|_{\alpha =0} = \langle -\partial_\mu \bar{\psi} \gamma^\mu \gamma^5 \psi - \frac{g^3}{8 \pi^2} Z^{\prime}_{\mu\nu} \Tilde{Z}^{\prime \mu \nu} + \frac{g^3C^{BBB}}{8 \pi^2 } \ Z^{\prime}_{\mu\nu} \Tilde{Z}^{\prime \mu \nu} \rangle \ ,
\end{equation}
where the last term comes from the St{\"u}ckelberg axion interaction with the Wess-Zumino term. Note that, in this example of the Wess-Zumino mechanism, the current conservation demands $C^{BBB}=1$. 

Before we describe our model, we should address the fact that the St{\"u}ckelberg axion described above is not a physical degree of freedom. This axion, in the St{\"u}ckelberg term, is {\it eaten} away to become the longitudinal component of the new gauge boson. On the other hand, in presence of other scalar fields in the model, this axion mixes with them, and integrating out the axion becomes non-trivial. To understand this, consider a scalar field ($\phi$) along with the axion field. The Lagrangian of the system is given by,
\begin{eqnarray}
    \mathcal{L} & = & -\frac{1}{4} Z^{\prime}_{\mu \nu} Z^{\prime \mu \nu} + L_{R_\xi} + \mathcal{L}_{scalar} \\ \nonumber
    \mathcal{L}_{scalar} &=& (D_\mu \phi)^{\dagger} (D^\mu \phi) -m_\phi^2 \phi^\dagger \phi+ \frac{1}{2}(\partial_\mu b + f_b Z^{\prime}_\mu)^2 \ ,
\end{eqnarray}
where $D_\mu = \partial_\mu - e_\phi g_{Z^{\prime}} Z^{\prime}_\mu$. Assuming that the new scalar field has a non-zero {\it vacuum expectation value}, $\phi = \frac{1}{\sqrt{2}}(v+\phi_1)e^{i \frac{\phi_2}{v}} $, the above Lagrangian, up to bilinear in fields, becomes,
\begin{eqnarray}
    \mathcal{L}_{scalar} & = & \dis \frac{1}{2} (\partial_\mu \phi_1)^2  - \frac{1}{2}m_\phi^2\phi_1^2 \nonumber  \\ 
    && \dis + \frac{1}{2}(\partial_\mu \phi_2)^2 + \frac{1}{2} (\partial_\mu b)^2 + \frac{1}{2}(f_b^2 + (e_\phi g_{Z^{\prime}} v)^2)Z^{\prime}_\mu Z^{\prime \mu} + Z^{\prime}_\mu \partial^\mu(f_b b+ e_\phi g_{Z^{\prime}} v \phi_2) \ .
\label{eq:scalar}
\end{eqnarray}
The gauge fixing term $L_{R_\xi}$ is generalised to include both the Goldstone bosons $b(x)$ and $\phi_2(x)$ as,\begin{equation}
L_{R_\xi} = -\frac{1}{2 \xi}(\partial_\mu Z^{\prime \mu} -\xi (f_b b+ e_\phi g_{Z^{\prime}} v \phi_2))^2 \ ,
\label{eq:generalisedRxi}
\end{equation}
so that the last term in Eq.\ref{eq:scalar} is cancelled. 
The first line in Eq.\ref{eq:scalar} is the Lagrangian for the real scalar field, and the second line contains the mixing of the Goldstone boson and the axion in the model. Upon diagonalising these terms by rotating to the physical basis ($G_1$ and $G_2$) of Goldstone bosons, given by,
\begin{eqnarray}
    G_1 & = & \dis \frac{1}{m_{Z^{\prime}}} (-f_b  \phi_2 + e_\phi g_{Z^{\prime}} v b) \ , \nonumber \\
    G_2 & = & \dis \frac{1}{m_{Z^{\prime}}} (e_\phi g_{Z^{\prime}} v \phi_2 + f_b b) \ ,
    \label{eq:physicalGoldstone}
\end{eqnarray}
 we get,
\begin{eqnarray}
    \mathcal{L}_{scalar} & = & \dis \frac{1}{2} (\partial_\mu \phi_1)^2  - \frac{1}{2}m_\phi^2\phi_1^2 \nonumber  \\ 
    && \dis + \frac{1}{2}(\partial_\mu G_1)^2 + \frac{1}{2} (\partial_\mu G_2)^2 + \frac{1}{2} m_{Z^{\prime}}^2 Z^{\prime}_\mu Z^{\prime \mu} + m_{Z^{\prime}} Z^{\prime}_\mu \partial^\mu G_2 \ ,
\label{eq:scalars-Goldstones}
\end{eqnarray}
where $m_{Z^{\prime}}^2=(f_b^2 + (e_\phi g_{Z^{\prime}} v)^2)$. 

From Eq.\ref{eq:generalisedRxi}, the generalised $R_\xi$ gauge now becomes,
\begin{eqnarray}
    \mathcal{L}_{R_\xi} = \frac{1}{\sqrt{\xi}}(\partial_\mu Z^{\prime \mu} - \xi m_{Z^{\prime}} G_2)^2.
\label{eq:goldstoneRxi}
\end{eqnarray}
The last term in Eq.\ref{eq:scalars-Goldstones} gets canceled by the mixing term in the above gauge fixing condition. Now it can be seen from Eq.\ref{eq:scalars-Goldstones} and Eq.\ref{eq:goldstoneRxi} that $G_1$ remains massless, while, the Goldstone boson $G_2$ receives a gauge dependent mass term. Thus $G_2$ will get {\it eaten} away by the $Z^{\prime}$ and becomes non-dynamical in the unitary gauge. Hence, we are left with one physical Goldstone boson $G_1$, namely the axi-Higgs.

While this Goldstone boson is physical, it remains massless due to the left over symmetry. Since, we need to make it massive, we need to break $U(1)^{\prime}$ explicitly. This can be achieved in the scalar Lagrangian~\cite{Coriano:2007fw} by including,
\begin{eqnarray}
    V_{\Slash{U}} &=& b_1 (\phi e^{-i e_\phi g_{Z^{\prime}} \frac{b}{f_b}}) + \lambda_1 (\phi e^{-i e_\phi g_{Z^{\prime}} \frac{b}{f_b}})^2 + 2 \lambda_2 (\phi^*\phi)(\phi e^{-i e_\phi g_{Z^{\prime}} \frac{b}{f_b}}) +  c.c. \nonumber \\
&=& b_1 v \ Cos\big(\frac{\phi_2}{v} -e_\phi g_{Z^{\prime}} \frac{b}{f_b}\big) + \lambda_1 v^2 Cos\big(2(\frac{\phi_2}{v}-e_\phi g_{Z^{\prime}} \frac{b}{f_b})\big) + 2\lambda_2 v^3 Cos\big(\frac{\phi_2}{v} -e_\phi g_{Z^{\prime}} \frac{b}{f_b}\big) \ ,
\end{eqnarray}
where we have used $\phi = \frac{1}{\sqrt{2}}(v+\phi_1)e^{i \frac{\phi_2}{v}} $.
Expanding this to the quadratic order, we get,
\begin{eqnarray}
V_{\Slash{U}} \supset -\frac{1}{2} c_{G_1} v^2 \begin{pmatrix} \phi_2 & b \end{pmatrix} \begin{pmatrix} 1 & - v  \frac{e_\phi g_{Z^{\prime}}}{f_b} \\
			 v  \frac{e_\phi g_{Z^{\prime}}}{f_b} & v^2 (\frac{e_\phi g_{Z^{\prime}}}{f_b})^2 \end{pmatrix} \begin{pmatrix} \phi_2 \\ b \end{pmatrix} \ .
\end{eqnarray}
In the above equation $c_{G_1} = 4(\frac{b_1}{v^3} + 4\frac{\lambda_1}{v^2} + 2 \frac{\lambda_2}{v})$. Upon diagonalising and using Eq.\ref{eq:physicalGoldstone}, the mass of the pseudo-Goldstone boson $G_1$ can be computed as,
\begin{eqnarray}
    m_{G_1}^2 = - \frac{1}{2} c_{G_1} v^2 (1+ (e_\phi g_{Z^{\prime}} v)^2/f_b^2) = -\frac{1}{2} c_{G_1}v^2 \frac{m_{Z^{\prime}}^2}{f_b^2}\ ,
    \label{eq:psuedoGoldstonemass}
\end{eqnarray}
Moreover, the scalar fields in the un-rotated basis are related to the physical basis as,
\begin{eqnarray}
    \phi_2 & =& \dis \frac{1}{m_{Z^{\prime}}}(-f_b G_1 +e_\phi g_{Z^{\prime}} v G_2) \ , \nonumber \\
    b & =& \dis \frac{1}{m_{Z^{\prime}}}(e_\phi g_{Z^{\prime}} v  G_1 +f_bG_2) \ .
    \label{eq:unrotatedphysical}
\end{eqnarray}
With the discussion of the ax-Higgs complete, in the next section, we will study how the anomalous interaction terms contribute to additional annihilation channels for fermionic dark matter via the axi-Higgs portal. 

\section{Anomalous $U(1)^{\prime}$ extension of the Standard Model}
\label{sec:model}

One of the first possible imminent discoveries at the {\it Large Hadron Collider} (LHC) would be a neutral gauge boson with a mass of $\mathcal{O}({\rm TeV})$. Various minimal extensions to the SM consider an enlarged gauge structure with abelian groups. These models predict resonances that could be searched at colliders, and the field has received attention due to the prediction of many such anomalous abelian gauge groups in D-brane models. On the other hand, such minimal extensions bring in gauge anomalies, which are either cancelled by a careful choice of gauge quantum numbers or a St{\"u}ckelberg axion, as described in the previous section. Here we are interested in studying an anomalous $U(1)^{\prime}$ extension to the Standard Model in a non-renormalisable effective field theory and the axi-Higgs present in the model. 
\begin{table}[!h]
\begin{center}
\begin{tabular}{|l|l|l|l|l|}
\hline
Field       & $SU(3)_{C}$  & $SU(2)_{W}$  & $U(1)_{Y}$    & $U(1)^{\prime}$        \\ \hline
$Q_{L}^{i}$ & 3            & 2            & 1/3           & $e_{q_L}^{i}= (e_{q_{L}}^1,e_{q_{L}}^2,e_{q_{L}}^3)$        \\
$u_{R}^{i}$ & 3            & 1            & 4/3           & $e_{u_R}^{i} =(e_{u_{R}},e_{c_{R}},e_{t_{R}})$        \\
$d_{R}^{i}$ & 3            & 1            & -2/3          & $e_{d_R}^{i} = (e_{d_{R}},e_{s_{R}},e_{b_{R}})$        \\
$L_{L}^{i}$ & 1            & 2            & -1            & $e^{i}_{\ell_L} = (e_{\ell_{L}}^1,e_{\ell_{L}}^2,e_{\ell_{L}}^3)$        \\
$e_{R}^{i}$ & 1            & 1            & -2            & $e^{i}_{\ell_R} = (e_{e_{R}},e_{\mu_{R}},e_{\tau_{R}})$        \\
$\chi_{L}$  & 1            & 1            & 0             & $e_{\chi_{L}}$ \\
$\chi_{R}$  & 1            & 1            & 0             & $e_{\chi_{R}}$ \\
$\phi$      & 1            & 1            & 0             & $e_{\phi}$     \\
H           & 1            & 2            & 1             & 0              \\ \hline
\end{tabular}
\caption {Particle content of the $U(1)^{\prime}$ model. The Weyl fermions $\chi_L$ and $\chi_R$ are the left-handed and right-handed dark matter particles and $\phi$ is the new scalar field that spontaneously breaks the $U(1)^{\prime}$ symmetry. i = 1, 2, 3 denotes the generation index.} \label{table1} 
\end{center}
\end{table}\\
To accommodate the massive vector boson, let's consider the extended Standard Model gauge group $SU(3)_{C} \times SU(2)_{W} \times U(1)_{Y} \times U(1)^{\prime}$, and matter content as given in Tab.\ref{table1}. The new abelian field being anomalous, leads to triangle diagrams with coefficients,
\begin{eqnarray}
    U(1)^{\prime}-U(1)^{\prime}-U(1)^{\prime} & : & \dis \mathcal{A}^{(0)} = \sum_f Q_f^3 \ , \nonumber \\
    U(1)^{\prime}-U(1)^{\prime}-U(1)_Y & : & \dis \mathcal{A}^{(1)} = \sum_f Q_f^2 Y_f \ , \nonumber \\
    U(1)^{\prime}-U(1)_Y-U(1)_Y & : & \dis \mathcal{A}^{(2)} = \sum_f Q_f Y_f^2 \ , \nonumber \\
    U(1)^{\prime}-SU(2)_W-SU(2)_W & : & \dis \mathcal{A}^{(3)} = \sum_f Q_f Tr[T_kT_k] \ , \nonumber \\
    U(1)^{\prime}-SU(3)_c-SU(3)_c & : & \dis \mathcal{A}^{(4)} = \sum_f Q_f Tr[\tau_k \tau_k] \ ,
\end{eqnarray}
where $f$ runs over all the fermions in Tab.\ref{table1}, ($Y_f$, $Q_f$) are the corresponding charges of ($U(1)_Y$, $U(1)^{\prime})$ abelian gauge groups and ($T_{k}$, $\tau_{k}$) are the generators of ($SU(2)_W$, $SU(3)_c$) respectively such that they satisfy $Tr[T_i T_j] = \frac{1}{2}\delta_{ij}=Tr[\tau_i \tau_j]$ and $Tr[T_i] = 0 =Tr[\tau_i]$. 
To cancel these anomalous contributions, we introduce the St{\"u}ckelberg axion interaction,
\begin{eqnarray}
    \mathcal{L}_{b} &=& \dis \frac{1}{48 \pi^{2}} \frac{g_{Z^{\prime}}}{f_{b}} b(x) \Big(C_{1} g_{Z^{\prime}}^{2} Z^{\prime}_{\mu\nu} \tilde{Z}^{\prime \mu\nu} + C_{2} g_{Y} g_{Z^{\prime}} Y_{\mu\nu} \tilde{Z}^{\prime \mu\nu} + C_{3} g_{Y}^{2} Y_{\mu\nu} \tilde{Y}^{\mu\nu} + C_{4} g_{W}^{2} W_{\mu\nu} \tilde{W}^{\mu\nu} + C_{5} g_{G}^{2} G_{\mu\nu} \tilde{G}^{\mu\nu}\Big) \ ,
    \label{eq:axionint}
\end{eqnarray}
Using the matter content in Tab.\ref{table1}, we then get,
\begin{eqnarray}
    C_{1} &=& \dis 6e_{q}^{3}-3e_{u}^{3}-3e_{d}^3+6e_{l}^3-3e_{e}^3+e_{\chi_{L}}^3-e_{\chi_{R}}^3 \ , \nonumber \\
    C_{2} &=& \dis 2e_{q}^{2}-4e_{u}^{2}+2e_{d}^2-6e_{l}^2+6e_{e}^2 \ , \nonumber \\
    C_{3} &=& \dis 2/3 e_{q}-16/3 e_{u}-4/3 e_{d}+6e_{l}-12e_{e} \ , \nonumber \\
    C_{4} &=& \dis 6e_{q}+6e_{l} \ , \nonumber \\
    C_{5} &=& \dis 18e_{q}-9e_{u}-9e_{d} \ .
    \label{eq:anomaly}
\end{eqnarray}
After the Electro-weak symmetry breaking, $SU(2)_W \times U(1)_Y \to U(1)_{em}$, the $W_\mu$ and $Y_\mu$ bosons mix to give rise to massive $Z_\mu$ and the massless photon $\gamma$. The resultant interaction Lagrangian of the axion becomes, 
\begin{eqnarray}
    \mathcal{L}_{b} &=& \dis  \frac{1}{12} g_{Z^{\prime}} b(x) \Big(g_{b\gamma\gamma} F_{\mu\nu} \tilde{F}^{\mu\nu} + g_{bZ\gamma} F_{\mu\nu} \tilde{Z}^{\mu\nu} + g_{bZZ} Z_{\mu\nu} \tilde{Z}^{\mu\nu} + g_{bWW} W^{+}_{\mu\nu} \tilde{W}^{-\mu\nu} \nonumber \\ 
    && \dis + g_{bZ^{\prime}Z^{\prime}} Z^{\prime}_{\mu\nu} \tilde{Z}^{\prime \mu\nu} + g_{b \gamma Z^{\prime}} F_{\mu\nu} \tilde{Z}^{\prime \mu\nu} + g_{bZZ^{\prime}} Z_{\mu\nu} \tilde{Z}^{\prime \mu\nu} + \frac{1}{4 \pi^{2}f_b}C_5 g_G^2 G_{\mu\nu} \tilde{G}^{\mu\nu} \Big) \ ,
    \label{eq:axionintafterEWSB}
\end{eqnarray}
where the couplings, in the broken phase is given by,
\begin{align}
    g_{b\gamma\gamma} &= \dis \frac{C_{3} g_{Y}^{2} c_{w}^{2} + C_{4} g_{W}^{2} s_{w}^{2}}{4 \pi^{2} f_{b}}  \ ,& g_{bZ\gamma} &= \dis \frac{C_{4} g_{W}^{2} s_{w} c_{w} - C_{3} g_{Y}^{2} s_{w} c_{w}}{2 \pi^{2} f_{b}}  \ ,\nonumber \\
    g_{bZZ} &= \dis \frac{C_{3} g_{y}^{2} s_{w}^{2} + C_{4} g_{w}^{2} c_{w}^{2}}{4 \pi^{2} f_{b}} \ , & g_{bWW}  &=  \dis \frac{C_{4} g_{w}^{2}}{2 \pi^{2} f_{b}}  \ ,\nonumber \\
    g_{bZ^{\prime}Z^{\prime}} &= \dis \frac{C_{1} g_{Z^{\prime}}^{2}}{4 \pi^{2} f_{b}}  \ , & g_{b\gamma Z^{\prime}} &= \dis \frac{C_{2} g_{Y} g_{Z^{\prime}} c_{w}}{4 \pi^{2} f_{b}}  \ , \nonumber \\
    g_{bZZ^{\prime}} &= \dis \frac{-C_{2} g_{Y} g_{Z^{\prime}} s_{w}}{4 \pi^{2} f_{b}} \ .
    \label{eq:axioncouplingafterEWSB}
\end{align}
In the above equation, $g_{Z^{\prime}} = \sqrt{\frac{m_{Z^{\prime}}^2-f_b^2}{{e_{\phi}^{2} v^2}}}$ and $s_{w}$ and $c_{w}$ are the sine and cosine of the Weinberg angle respectively.

With the extended Standard Model, the Lagrangian density of the matter content becomes,
\begin{eqnarray}
    \mathcal{L}_{fermion} & = & \dis \sum_{i=1}^3 \Big( \overline{Q}_L^{i} i \Slash{D} Q_L^{i} + \overline{u}_R^{i} i \Slash{D} u_R^{i} + \overline{d}_R^{i} i \Slash{D} d_R^{i} + \overline{L}_L^{i} i \Slash{D} L_L^{i} + \overline{e}_R^{i} i \Slash{D} e_R^{i}\Big) \nonumber \\
    && \dis + \overline{\chi}_L i \Slash{D} \chi_L + \overline{\chi}_R i \Slash{D} \chi_R + \mathcal{L}_{Yukawa} \ ,
\end{eqnarray}
where $\Slash{D}=\gamma^\mu D_\mu$, with $D_\mu$ representing the covariant derivatives containing the respective gauge interactions. The Yukawa term incorporates the scalar-fermion bilinear interactions given by,
\begin{eqnarray}
    \mathcal{L}_{Yukawa} &=& \dis -\lambda^{ij}_{u} (\frac{\phi^{\dagger}}{\Lambda})^{r_{u ij}} \overline{Q}^{i}_{L} \tilde{H} u^{j}_{R} - \lambda^{ij}_{d} (\frac{\phi^{\dagger}}{\Lambda})^{r_{d ij}} \overline{Q}^{i}_{L} H d^{j}_{R} - \lambda^{ij}_{\ell}(\frac{\phi^{\dagger}}{\Lambda})^{n_{ij}} \overline{L}^{i}_L H e^{j}_R - \lambda_\chi \phi^{\dagger} (\frac{\phi^{\dagger}}{\Lambda})^{n_{4}} \overline{\chi}_{L} \chi_{R} + \text{h.c.} \ ,
    \label{eq:LYukawa}
\end{eqnarray}
where, $\tilde{H} = i\sigma_2 H$ and $r_{u ij} = e_{q_{L}}^{i}-e_{u_{R}}^{j}$, $r_{d ij} = e_{q_{L}}^{i}-e_{d_{R}}^{j}$, $n_{ij} = e_{\ell_{L}}^{i}-e_{\ell_{R}}^{j}$ and $n_4 = e_{\chi_{L}}-e_{\chi_{R}}+e_{\phi}$.

Upon expanding about the {\it vacuum expectation value} of $\phi$, the Goldstone boson $\phi_2$ appears as phase and can be rotated away from the Yukawa terms by redefining the fermions. Instead, they re-emerge as interaction terms from the kinetic term of the fermions. Working in the basis in which the Electro-weak symmetry is broken, the $\phi_{2}$-fermion interaction becomes,
\begin{eqnarray}
    \mathcal{L}_{Goldstone-fermion} & = & \dis \frac{r_{uii}}{2v}(m_{u^i}) (\phi_2 \overline{u}^i\gamma^5 u^i) + \frac{r_{dii}}{2v}(m_{d^i})(\phi_2\overline{d}^i\gamma^5 d^i) + \frac{n_{ii}}{2v}(m_{e^i} )(\phi_2 \overline{e}^i\gamma^5 e^i) + \frac{n_4+1}{2v}(m_{\chi})( \phi_2 \overline{\chi} \gamma^5 \chi) \ . \nonumber \\
    \label{eq:Golstonefermion}
\end{eqnarray}
where $u^{i}$, $d^{i}$, $e^{i}$ and $\chi^{i}$ are Dirac fermions consisting of both left and right spinors.

Since the scalar field, $\phi$, transforms non-trivially under $U(1)^{\prime}$, its Goldstone mode mix with the St{\"u}ckelberg axion as shown in the previous section. The SM Higgs field, on the other hand, does not mix with these axions since they are not charged under $U(1)^{\prime}$. With this mixing, the physical Goldstone boson becomes a mixture of the axion ($b$) and the Goldstone ($\phi_2$) as given in Eq.\ref{eq:physicalGoldstone} with mass given in Eq.\ref{eq:psuedoGoldstonemass}.
The interaction of the physical pseudo-Goldstone boson (axi-Higgs), $G_1$, could be found by re-writing Eq.\ref{eq:axionint} and Eq.\ref{eq:Golstonefermion} using Eq.\ref{eq:physicalGoldstone}. In the gauge $\xi \to \infty$, the $G_2$ becomes non-dynamical and using Eq.\ref{eq:unrotatedphysical} we get,
\begin{eqnarray}
    \mathcal{L}_{G_1BB} &=& \dis  \frac{1}{12} g_{Z^{\prime}} \frac{e_\phi g_{Z^{\prime}} v}{m_{Z^{\prime}}}  G_1 \Big(g_{b\gamma\gamma} F_{\mu\nu} \tilde{F}^{\mu\nu} + g_{bZ\gamma} F_{\mu\nu} \tilde{Z}^{\mu\nu} + g_{bZZ} Z_{\mu\nu} \tilde{Z}^{\mu\nu} + g_{bWW} W^{+}_{\mu\nu} \tilde{W}^{-\mu\nu} \nonumber \\ 
    && \dis + g_{bZ^{\prime}Z^{\prime}} Z^{\prime}_{\mu\nu} \tilde{Z}^{\prime \mu\nu} + g_{b \gamma Z^{\prime}} F_{\mu\nu} \tilde{Z}^{\prime \mu\nu} + g_{bZZ^{\prime}} Z_{\mu\nu} \tilde{Z}^{\prime \mu\nu} + \frac{1}{4 \pi^{2}f_b}C_5 g_G^2 G_{\mu\nu} \tilde{G}^{\mu\nu} \Big) \ ,
    \label{eq:axiHiggsafterEWSB}
\end{eqnarray}
and 
\begin{eqnarray}
     \mathcal{L}_{G_1fermion} & = & \dis - \frac{r_{uii}}{2v} \frac{f_b m_{u^i}}{m_{Z^{\prime}}}( G_1\overline{u}^i\gamma^5 u^i) - \frac{r_{dii}}{2v}\frac{f_b m_{d^i}}{m_{Z^{\prime}}}( G_1 \overline{d}^i\gamma^5 d^i) \nonumber \\
     && \dis - \frac{n_{ii}}{2v}\frac{f_b m_{e^i}}{m_{Z^{\prime}}}( G_1\overline{e}^i\gamma^5 e^i) - \frac{n_4+1}{2v}\frac{f_b m_{\chi}}{m_{Z^{\prime}}}(G_1 \overline{\chi} \gamma^5\chi) \ .
    \label{eq:axiHiggsfermion}
\end{eqnarray}
Note that, here we have re-named the Lagrangian densities $\mathcal{L}_b$ and $\mathcal{L}_{Goldstone-fermion}$ as $\mathcal{L}_{G_1 BB}$ and $\mathcal{L}_{G_{1}fermion}$ respectively.

\begin{figure}[!h]
\hspace{-1cm}
  \includegraphics[width=20cm,height=8cm]{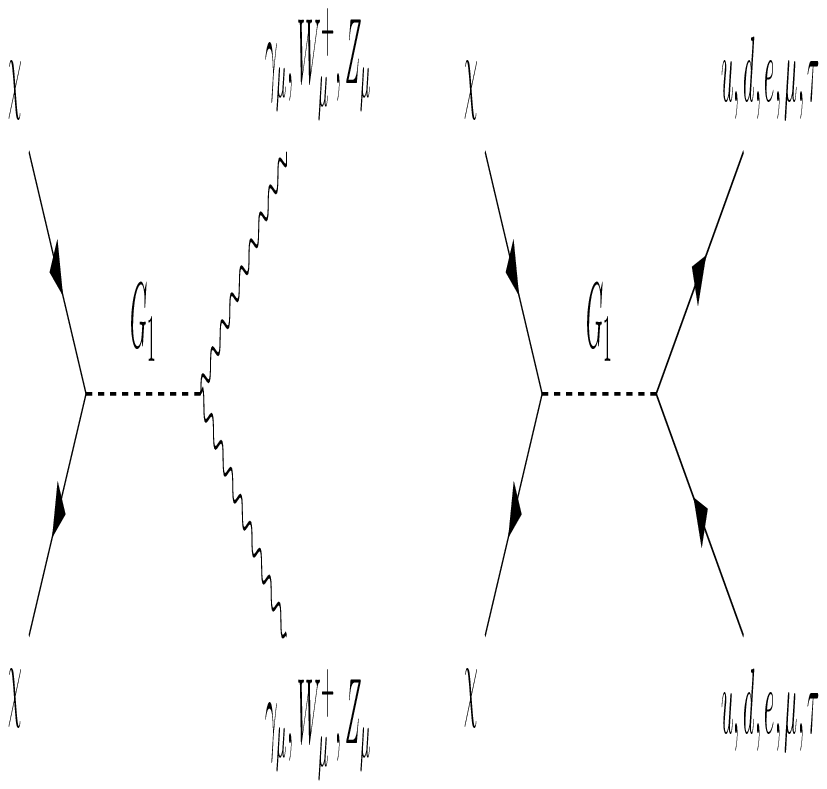}
  \vspace{-1.5cm}
\caption{Diagrams that contribute to dark matter annihilation.}
\label{fig:DMannihilation}
\end{figure}

This article aims to show that the Wess-Zumino interaction term of axi-Higgs, in Eq.\ref{eq:axiHiggsafterEWSB}, and axi-Higgs fermion interaction in Eq.\ref{eq:axiHiggsfermion} together foster an efficient portal for dark matter annihilation as shown in Fig.\ref{fig:DMannihilation}. Though dark matter can also annihilate via $Z^{\prime}$ mediation, this cross-section is suppressed due to the $\mathcal{O}(\text{TeV})$ mass of $Z^{\prime}$. In this model, we consider the Dirac fermion $\chi$ to be the dark matter candidate and, in the next section, find the parameter space in which it satisfies the observed relic density of the Universe. Though this axi-Higgs portal exists in various anomalous $U(1)^{\prime}$ extension of the Standard Model, for clarity of reading, we choose two simple models. 

For simplicity, here we consider anomalous $U^{\prime}_{e_\mu L_\mu- e_\tau L_\tau}$ and anomalous $U^{\prime}_{\gamma_5(L_e+L_\mu+L_\tau)}$ models, in which axi-Higgs do not couple to quarks and gluons. Thus the parameter space constrained by LHC and gauge invariance opens up for massive axi-Higgs. Moreover, we choose a region in which the charged leptonic flavour violations also are suppressed. The study of generic anomalous $U(1)^{\prime}$ with all phenomenological considerations is reserved for a future communication. 
    
\subsection{Anomalous $U^{\prime}_{e_\mu L_\mu- e_\tau L_\tau}$}
\label{sec:firstmodel}
\begin{table}[H]
\begin{center}
\begin{tabular}{|l|lllllllll|}
\hline
Field        & $Q_{L}^{i}$ & $u_{R}^{i}$ & $d_{R}^{i}$ & $L_{L}^{i}$ & $e_{R}^{i}$ & $\chi_{L}$ & $\chi_{R}$ & $\phi$ & H  \\ \hline
$U(1)^{\prime}$ charge & 0   & 0   & 0   & $(0,e_\mu,-e_\tau)$  & $(0,e_\mu,-e_\tau)$ & 1/2 & -1/2  & -1  & 0 \\ \hline
\end{tabular}
\caption {Particle content of the anomalous $U(1)^{\prime}_{e_\mu L_\mu -e_\tau L_\tau}$ model.}
\label{table:LmuLtau} 
\end{center}
\end{table}

For the given set of charges for the fermions, using Eq.\ref{eq:anomaly}, the anomaly cancellation conditions become,
\begin{equation}
    C_{1} = e_{\mu}^3-e_{\tau}^3 + 1/4, \hspace{9pt} C_{2} = 0, \hspace{9pt} C_{3} = -2e_{\mu}+2e_{\tau}, \hspace{9pt} C_{4} =2 e_{\mu}-2 e_{\tau}, \hspace{9pt} C_{5} = 0
    \label{eq:anomalycoeffmutau}
\end{equation}
And, the Yukawa interactions in Eq.\ref{eq:LYukawa} now becomes,
\begin{equation}
    \mathcal{L}_{Yukawa} = -\lambda_{q} \overline{Q}_{L} \tilde{H} u_{R} - \lambda_{d} \overline{Q}_{L} H d_{R} - \lambda_{e}\overline{L}^{1}_{L} H e^{1}_{R} - \lambda_{\mu} \overline{L}^{2}_L H e^{2}_{R} - \lambda_{\tau} \overline{L}^{3}_L H e^{3}_{R} - \lambda_{mix} \Big( \frac{\phi^{\dagger}}{\Lambda}\Big)^{e}\overline{L}^{2}_L H e^{3}_{R} - \phi^\dagger \bar{\chi_{L}} \chi_{R} + h.c. \ ,
    \label{eq:mutauYukawa}
\end{equation}
where $e = e_\mu+e_\tau$.\\
In the above, we have the freedom to choose `$e$', $\lambda_{mix}$ and $\frac{\langle \phi \rangle}{\Lambda}$ such that the charged lepton mixing is suppressed, making the model free from charged lepton flavour violations. For completeness, here we take $e_{\mu}=2$ and $e_{\tau}=1$. In this scenario, $g_{b\gamma \gamma}$ vanishes, but though it can be generated at higher loop level, here we will not consider that. Using Eq.\ref{eq:mutauYukawa} and Eq.\ref{eq:anomalycoeffmutau}, the axi-Higgs couplings given in Eq.\ref{eq:axiHiggsafterEWSB} and Eq.\ref{eq:axiHiggsfermion} becomes,
\begin{eqnarray}
    \mathcal{L}_{G_1BB} &=& \dis  - \frac{1}{12} g_{Z^{\prime}} \frac{g_{Z^{\prime}} v}{m_{Z^{\prime}}}  G_1 \Big(g_{bZ\gamma} F_{\mu\nu} \tilde{Z}^{\mu\nu} + g_{bZZ} Z_{\mu\nu} \tilde{Z}^{\mu\nu} + g_{bWW} W^{+}_{\mu\nu} \tilde{W}^{-\mu\nu} \nonumber \\ 
    && \dis + g_{bZ^{\prime}Z^{\prime}} Z^{\prime}_{\mu\nu} \tilde{Z}^{\prime \mu\nu} + g_{b \gamma Z^{\prime}} F_{\mu\nu} \tilde{Z}^{\prime \mu\nu} + g_{bZZ^{\prime}} Z_{\mu\nu} \tilde{Z}^{\prime \mu\nu} \Big) \ ,
    \label{eq:axiHiggscouplingmutauanomaly}
\end{eqnarray}
and 
\begin{eqnarray}
     \mathcal{L}_{G_1fermion} & = & \dis - \frac{1}{2v}\frac{f_b m_{\chi}}{m_{Z^{\prime}}}(G_1\overline{\chi} \gamma^5 \chi) \ .
    \label{eq:axiHiggsmutau}
\end{eqnarray}
These interaction terms lead to annihilation channels for the dark matter $\chi \overline{\chi} \to (WW,ZZ,Z\gamma)$ as shown in the left diagram in Fig.\ref{fig:DMannihilation}.

\subsection{Anomalous $U(1)^{\prime}_{\gamma_5(L_e+L_\mu+L_\tau)}$}
\label{sec:secondmodel}
The charges of the fermion and scalar fields under the $U(1)^{\prime}_{\gamma_5(L_e+L_\mu+L_\tau)}$ symmetry are given in Tab.\ref{table:LR}.\\

\begin{table}[H]
\begin{center}
\begin{tabular}{|l|lllllllll|}
\hline
Field        & $Q_{L}^{i}$ & $u_{R}^{i}$ & $d_{R}^{i}$ & $L_{L}^{i}$ & $e_{R}^{i}$ & $\chi_{L}$ & $\chi_{R}$ & $\phi$ & H  \\ \hline
$U(1)^{\prime}$ charge & 0   & 0   & 0   & (e,e,e)  & (-e,-e,-e) & 1/2 & -1/2  & -1  & 0 \\ \hline
\end{tabular}
\caption {Particle content of the anomalous $U(1)^{\prime}_{\gamma_5(L_e+L_\mu+L_\tau)}$  model.}
\label{table:LR} 
\end{center}
\end{table}
Here, the scalar field interactions are given by the Lagrangian,
\begin{equation}
    \mathcal{L}_{H} = -\lambda_{q} \Bar{Q_{L}} \tilde{H} u_{R} - \lambda_{d} \bar{Q_{L}} H d_{R} + \lambda_{e}^{ij}(\frac{\phi^\dagger}{\Lambda})^{2} \overline{L}_{L}^{i} H e_{R}^{j}+ \phi^\dagger \bar{\chi_{L}} \chi_{R} \ .
    \label{eq:emutauYukawa}
\end{equation}
From Eq.\ref{eq:anomaly}, for the given set of charges for the fermions, the cancellation of the $U(1)^{\prime}$ anomalies demands,

\begin{equation}
    C_{1} = 9e^3+1/4 , \hspace{9pt} C_{2} = 0, \hspace{9pt} C_{3} = 18e, \hspace{9pt} C_{4} =6e, \hspace{9pt} C_{5} = 0
    \label{eq:anomalycoeffemutau}
\end{equation}
Like before, for simplicity, we will consider $e=1$.
Tab.\ref{table:LR} ensures that the left-handed and right-handed leptons are charged uniformly under $U(1)^{\prime}$, hence not leading to lepton flavour violating decays. For two different benchmark $Z^{\prime}$ boson masses of 2.5 TeV and 3.5 TeV, we study the different co-annihilation channels of dark matter to standard model particles via the axi-Higgs portal, for dark matter mass in the range of 70 GeV to 460 GeV. 

Using Eq.\ref{eq:emutauYukawa} and Eq.\ref{eq:anomalycoeffemutau}, the axi-Higgs couplings given in Eq.\ref{eq:axiHiggsafterEWSB} and Eq.\ref{eq:axiHiggsfermion} becomes,
\begin{eqnarray}
    \mathcal{L}_{G_1BB} &=& \dis  - \frac{1}{12} g_{Z^{\prime}} \frac{g_{Z^{\prime}} v}{m_{Z^{\prime}}}  G_1 \Big(g_{b\gamma\gamma} F_{\mu\nu} \tilde{F}^{\mu\nu} + g_{bZ\gamma} F_{\mu\nu} \tilde{Z}^{\mu\nu} + g_{bZZ} Z_{\mu\nu} \tilde{Z}^{\mu\nu} + g_{bWW} W^{+}_{\mu\nu} \tilde{W}^{-\mu\nu} \nonumber \\ 
    && \dis + g_{bZ^{\prime}Z^{\prime}} Z^{\prime}_{\mu\nu} \tilde{Z}^{\prime \mu\nu} + g_{b \gamma Z^{\prime}} F_{\mu\nu} \tilde{Z}^{\prime \mu\nu} + g_{bZZ^{\prime}} Z_{\mu\nu} \tilde{Z}^{\prime \mu\nu} \Big) \ ,
    \label{eq:axiHiggscouplingallanomaly}
\end{eqnarray}
and 
\begin{eqnarray}
       \mathcal{L}_{G_1fermion} & = & \dis - \frac{1}{v}\frac{f_b m_{e^i}}{m_{Z^{\prime}}}( G_1 \overline{e}^i\gamma^5 e^i) - \frac{1}{2v}\frac{f_b m_{\chi}}{m_{Z^{\prime}}}( G_1 \overline{\chi} \gamma^5 \chi) \ .
    \label{eq:axiHiggsall}
\end{eqnarray}
These interaction terms lead to annihilation channels for the dark matter $\chi \overline{ \chi} \to (WW,ZZ,\gamma\gamma, e^{i}e^{i})$ as shown in Fig.\ref{fig:DMannihilation}.

\section{Relic Density}
\label{sec:relic}
Since neither the particle content nor the interactions of dark matter is known, we do not know the origin of the observed relic density in our universe. Nevertheless, there are two popular mechanisms that explain the observed abundance of dark matter - the freeze-out and the freeze-in mechanisms. In the thermal freeze-out mechanism, the dark matter is assumed to be initially in thermal equilibrium with the rest of the cosmic plasma. As the universe cooled and the temperature dropped below $\sim$ 0.05 times the mass of the dark matter, the Hubble expansion rate (the expansion rate of the Universe) exceeded the annihilation rate and they decoupled from the visible sector. As mentioned in Sec.\ref{sec:intro}, in this article we consider a Weakly Interacting Massive Particle (WIMP) freeze-out dark matter. Hence, the DM is assumed to freeze-out when the DM annihilation rate, $\Gamma$, is of the order of the Hubble rate $H$~\cite{Lisanti_2016} i.e., 
\begin{equation}
    \Gamma = n_{eq}\braket{\sigma v} \sim H .
\end{equation} 
Here, $n_{eq}$ is the dark matter number density when it is in thermal equilibrium with the Standard Model particles and the photon bath and $\braket{\sigma v}$ is the velocity-averaged annihilation cross section.
%When $\Gamma \gg$ H, DM readily annihilates and equilibrium can be maintained between the DM and photon bath. Instead, if $\Gamma \ll$ H, the DM particles fall out of equilibrium due to very low annihilation probability. Since our dark matter candidate ($\chi$), given in Tab.\ref{table1}, is a Cold dark matter (CDM) it is non-relativistic and collisionless, $n \sim T^{3/2} e^{-m_{\chi}/T}$, where T and $m_{\chi}$ are the temperature and mass of the DM respectively. Then, its phase-space distribution, $f(x,v)$, is given by the simplified Boltzmann equation,
%\begin{equation}
%    \frac{\partial f}{\partial t} + \dot{x} \frac{\partial f}{\partial x} + \dot{v} \frac{\partial f}{\partial v} = 0 
%\end{equation}

In our model, the dark matter number density today is calculated by the evolution of the $2 \to 2$ inelastic scattering processes shown in Fig.\ref{fig:DMannihilation} using the Boltzmann equation,
\begin{equation}
    \frac{dn}{dt} + 3Hn = - \braket{\sigma v} (n^2 - n_{eq}^2) ,
    \label{eq:boltzmann}
\end{equation}
where $n$ is the number density of DM. 
%This number density decreases with the expansion of the Universe and hence it is useful to define the abundance as Y = $\frac{n}{s}$, where $s$ is the total entropy density of the universe and $\dot{s} = -3sH$.  For a change of variable from time to temperature, the dimensionless quantity x is defined as x = $\frac{m_{\chi}}{T}$, and 
This relation is guided by the thermally averaged cross-section which is defined as~\cite{GONDOLO1991145},
\begin{equation}
    \braket{\sigma v}(T) = \frac{1}{8m_{\chi}^{4} T K_{2}^{2} (m_{\chi}/T)} \int_{4 m_{\chi}^{2}}^{\infty} d\tilde{s}  \sigma(\tilde{s}) (\tilde{s} - 4m_{\chi}^{2})\sqrt{\tilde{s}}K_1 \frac{\sqrt{\tilde{s}}}{T}
\end{equation}
where $K_i$ denotes the modified Bessel function of order $i$ and $\tilde{s}$ = $(p_1 + p_2)^2$, $p_1$ and $p_2$ being the four momenta of the annihilating dark matter particles. 
%Here all the densities and Møller velocity are defined in the cosmic comoving frame. 
The annihilation cross-sections $\sigma$ are computed for each process shown in Fig.\ref{fig:DMannihilation} , and they become,
\begin{eqnarray}
    \sigma(\chi \overline{\chi} \to W^+ W^-) &=& \dis (\frac{-f_{b} m_{\chi}}{2v m_{Z^{\prime}}})^2\frac{g_{bWW}^2 \tilde{s} (\tilde{s}-4m_W^2)}{16 \pi \Big((m_{G_1}^2-\tilde{s})^2+ m_{G_1}^2 \Gamma_{G_1}^2\Big)} \nonumber \\
    \sigma(\chi \overline{\chi} \to ZZ) &=& \dis (\frac{-f_{b} m_{\chi}}{2v m_{Z^{\prime}}})^2\frac{g_{bZZ}^2 \tilde{s} (\tilde{s}-4m_Z^2)}{16 \pi \Big((m_{G_1}^2-\tilde{s})^2+ m_{G_1}^2 \Gamma_{G_1}^2 \Big)} \nonumber \\
    \sigma(\chi\overline{\chi}\to \gamma \gamma) &=& \dis (\frac{-f_{b} m_{\chi}}{2v m_{Z^{\prime}}})^2\frac{g_{b\gamma \gamma}^2 \tilde{s}^{2}}{16 \pi \Big((m_{G_1}^2-\tilde{s})^2+ m_{G_1}^2 \Gamma_{G_1}^2\Big)} \nonumber \\
    \sigma(\chi \overline{\chi} \to \gamma Z) &=& \dis (\frac{-f_{b} m_{\chi}}{2v m_{Z^{\prime}}})^2\frac{g_{b\gamma Z}^2 (\tilde{s}+m_Z^2)^2}{16 \pi \Big((m_{G_1}^2-\tilde{s})^2+ m_{G_1}^2 \Gamma_{G_1}^2\Big)} \nonumber \\
    \sigma(\chi \overline{\chi}\to e^i\bar{e}^i) &=& \dis (\frac{-f_{b} m_{\chi}}{2v m_{Z^{\prime}}})^2(\frac{-f_{b} m_{e^i}}{v m_{Z^{\prime}}})^2\frac{\tilde{s}}{16 \pi \Big((m_{G_1}^2-\tilde{s})^2+ m_{G_1}^2 \Gamma_{G_1}^2\Big)} \ .
    \label{eq:crosssection}
\end{eqnarray}
Rewriting Eq.\ref{eq:boltzmann} with the abundance $Y = \frac{n}{s}$, where $s$ is the total entropy density of the universe, the evolution of the DM abundance with the evolution of the universe, with the rescaled time variable $x=m/T$, is described by,
\begin{equation}
    \frac{dY}{dx} = - \frac{xs \braket{\sigma v}}{H(m)} (Y^2 - Y_{eq}^{2}) \ .
\label{eq:abundance}
\end{equation}  
Using the solution for DM abundance today from the above equation, for a CDM in the freeze-out scenario, the relic density is given by,
\begin{equation}
     \Omega h^{2} = \frac{m_{\chi} Y_{\infty} s_{0} h^{2}}{\rho_{c}} \sim \frac{10^{-26} cm^3 s^{-1}}{\braket{\sigma v} \ ,}
     \label{eq:relic}
\end{equation}
where $Y_{\infty}$ and $s_{0}$ are the present-day DM yield and entropy density respectively, and $\rho_{c}$ is the critical density.
%Using $s_{0}$ = 2970 $cm^{-3}$ and $\rho_{c}$ = 1.054 $\times 10^{-5} h^{2}$ GeV $cm^{-3}$, 
Comparing this with critical dark matter density observed by the {\it Planck} collaboration~\cite{Planck:2018vyg},
\begin{equation}
    \Omega h^{2}_{critical} = 0.120 \pm 0.001 \ , 
\label{eq:criticalrelic}
\end{equation}
constraints the parameter space allowed in our model.

We use \texttt{micrOMEGAs 5.3.41}~\cite{Pukhov:2022ofh} for calculating the relic density for different parameters for the two models we described in subsections \ref{sec:firstmodel} and \ref{sec:secondmodel}. Since the dark matter does not interact with quarks or gluons in either of these models, {\it Large Hadron Collider} and direct detection bounds are relaxed.

%From collider experiments~\cite{Alonso-Alvarez:2018irt}, we set $m_{Z^{\prime}}$ $>$ 1.5 TeV. 
To compute the relic density, we consider two benchmark points for the mass of the $Z^{\prime}$ boson, $m_{Z^{\prime}}$ = 2.5 TeV and $m_{Z^{\prime}}$ = 3.5 TeV. Using the relation $m_{Z^{\prime}}^2 = (f_b^2+(e_\phi g_{Z^{\prime}}v)^2)$, for a given $g_{Z^{\prime}}$, $e_\phi$ and $v$, we get an upper limit on the value of the symmetry-breaking scale $f_b$ and hence the mass of the axi-Higgs $m_{G_{1}}$.

In the $U^{\prime}_{e_\mu L_\mu -e_\tau L_\tau}$ model, since the charge distribution prohibits DM-fermion coupling, the dark matter can annihilate only to the gauge bosons governed by the interaction Lagrangian given in Eq.\ref{eq:axiHiggscouplingmutauanomaly}. Hence, for the lower mass range ($m_{DM} \lesssim$ 45 GeV) dark matter relic density becomes higher than the observed critical value.

\begin{figure}[!h]
\hspace{-0.5cm}
    \begin{subfigure}{0.3\linewidth}
        \includegraphics[width=9cm,height=7cm]{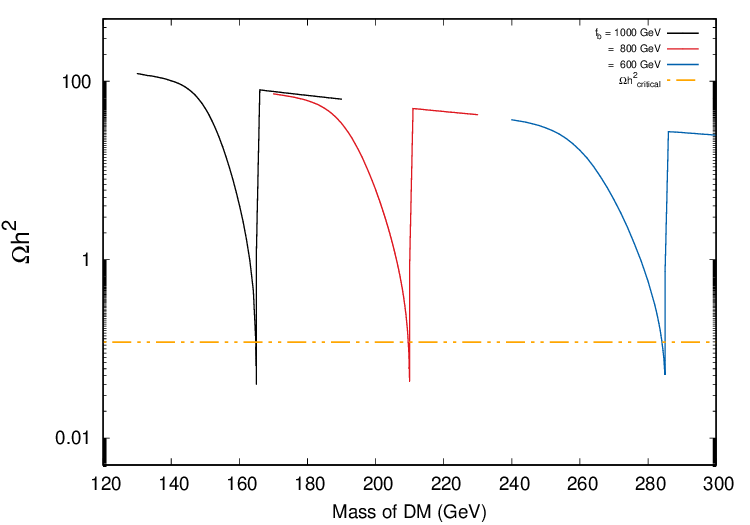}
              \caption{}
              \label{fig:relicLRmutau2500}
    \end{subfigure}
    \hspace{4.5cm}
    \begin{subfigure}{0.3\linewidth}
        \includegraphics[width=9cm,height=7cm]{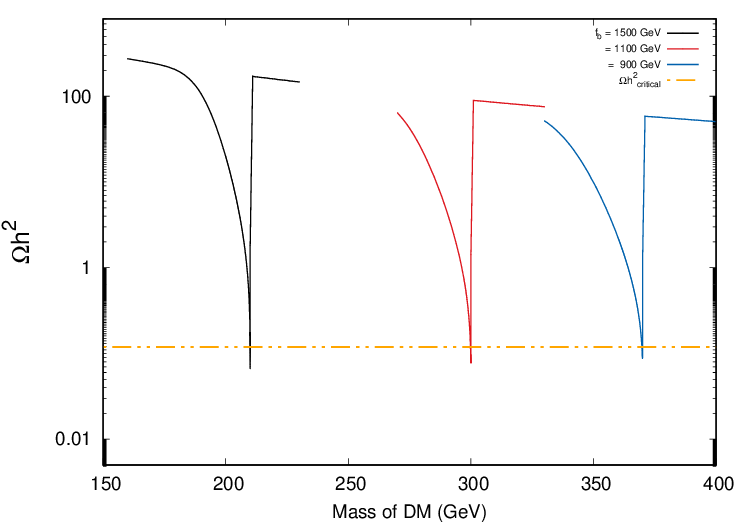}
              \caption{}
               \label{fig:relicLRmutau3500}
    \end{subfigure}
\caption{Plot showing the variation of relic density as a function of DM mass for different values of $f_b$, keeping $m_{Z^{\prime}}$ = 2500 GeV (left), 3500 GeV (right), $c_{G_{1}}$ = - 0.001 and $g_{Z^{\prime}} \sim 0.4$ for the $U^{\prime}_{L_\mu - L_\tau}$ model. Mass of the axi-higgs is related to these parameters as $m_{G1} = \frac{m_{Z^{\prime}} \sqrt{c_{G1} (m_{Z^{\prime}}^2-f_{b}^2)}}{\sqrt{2} g_{Z^{\prime}} f_{b}}$.}
\label{fig:relicLRmutau}
\end{figure}

\begin{figure}[h]
    \begin{subfigure}{0.3\linewidth}
        \includegraphics[width=9cm,height=8cm]{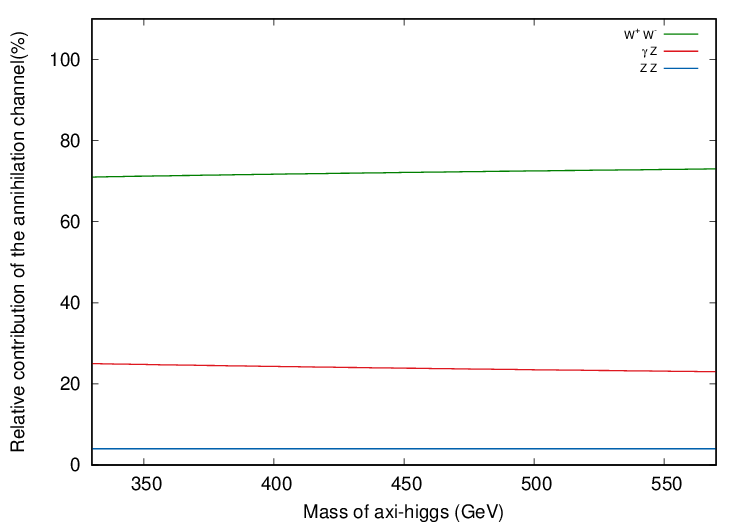}
        \caption{}
        \label{fig:channels2500LRmutau}
    \end{subfigure}
    \hspace{4cm}
    \begin{subfigure}{0.3\linewidth}
        \includegraphics[width=9cm,height=8cm]{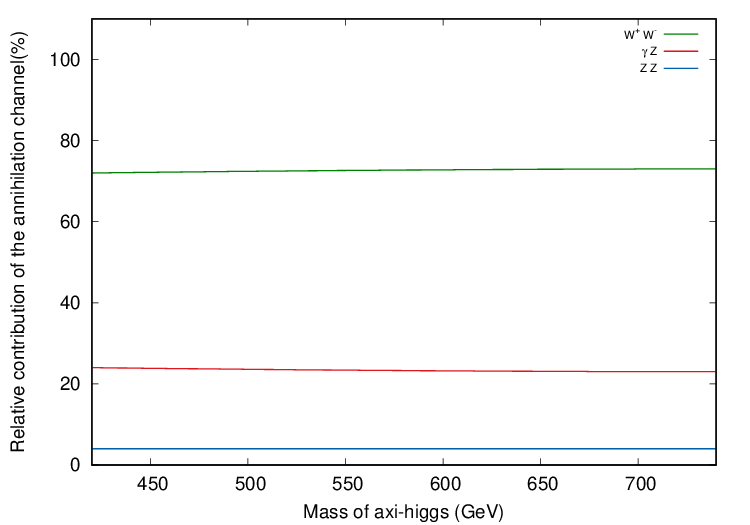}
        \caption{}
        \label{fig:channels3500LRmutau}
    \end{subfigure}
\caption{Relative contribution of the different annihilation channels vs mass of mediator plot for which the calculated relic density matches $\Omega h^{2}_{critical}$, for $m_{Z^{\prime}}$ = 2500 GeV (left), 3500 GeV (right), $c_{G_{1}}$ = - 0.001  and $g_{Z^{\prime}} \sim 0.4$ for the $U(1)^{\prime}_{L_{\mu}-L_{\tau}}$ model}
    \label{fig:channelsLRmutau}
  \end{figure}

\begin{figure}[!h]
\hspace{-0.5cm}
    \begin{subfigure}{0.3\linewidth}
        \includegraphics[width=9cm,height=7cm]{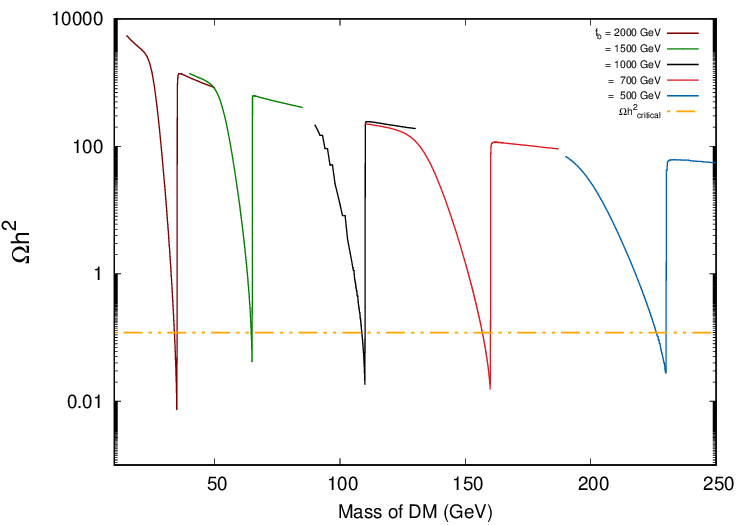}
              \caption{}
		\label{fig:relicLR}
    \end{subfigure}
    \hspace{4.5cm}
    \begin{subfigure}{0.3\linewidth}
        \includegraphics[width=9cm,height=7cm]{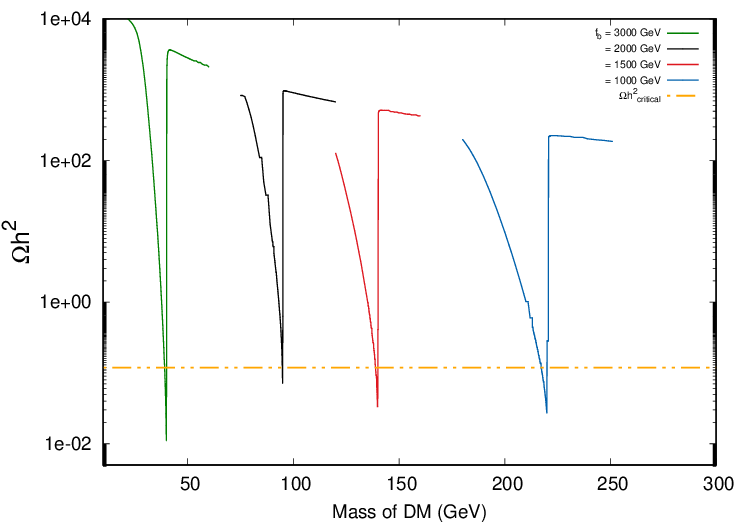}
              \caption{}
		\label{fig:omega3500}
    \end{subfigure}
\caption{Plot showing the variation of relic density as a function of DM mass for different values of $f_b$, keeping $m_{Z^{\prime}}$ = 2500 GeV (left), 3500 GeV (right), $c_{G_{1}}$ = - 0.001 and $g_{Z^{\prime}} \sim 0.6$ for the $U^{\prime}_{L_e + L_\mu + L_\tau}$ model. Mass of the axi-higgs is related to these parameters as $m_{G1} = \frac{m_{Z^{\prime}} \sqrt{c_{G1} (m_{Z^{\prime}}^2-f_{b}^2)}}{\sqrt{2} g_{Z^{\prime}} f_{b}}$.}
\label{fig:omegaall}
\end{figure}

\begin{figure}[!h]
    \begin{subfigure}{0.3\linewidth}
        \includegraphics[width=9cm,height=8cm]{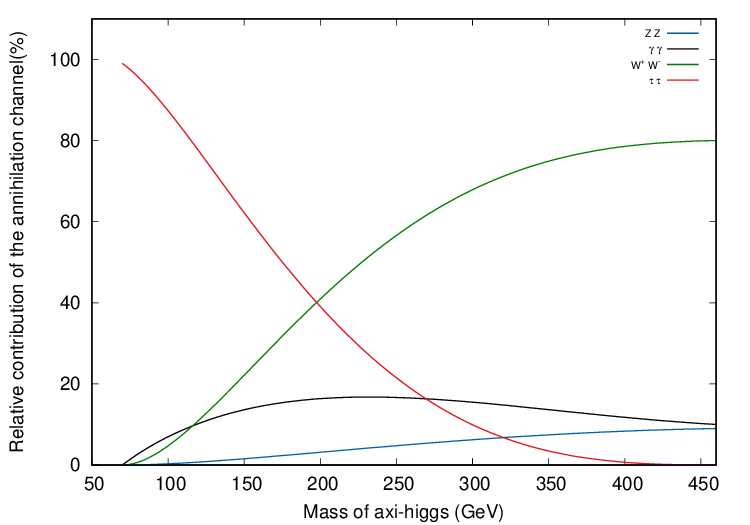}
        \caption{}
		\label{fig:channels}
    \end{subfigure}
    \hspace{4cm}
    \begin{subfigure}{0.3\linewidth}
        \includegraphics[width=9cm,height=8cm]{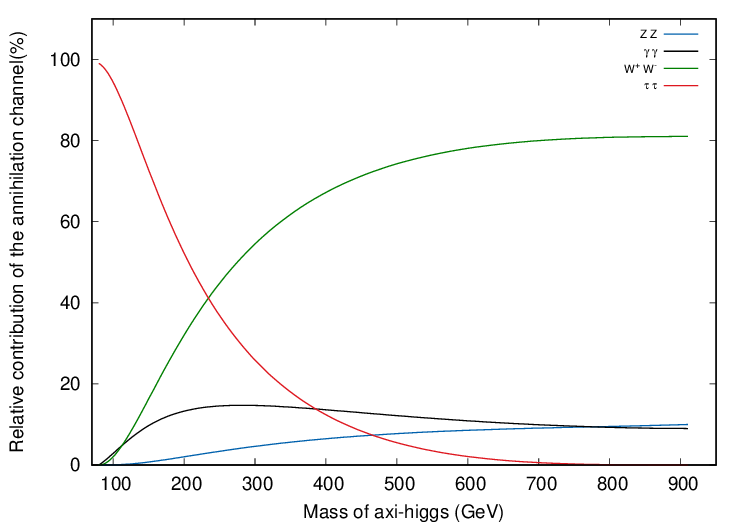}
        \caption{}
	    \label{fig:channels3500}
    \end{subfigure}
\caption{Relative contribution of the different annihilation channels vs mass of mediator plot for which the calculated relic density matches $\Omega h^{2}_{critical}$, for $m_{Z^{\prime}}$ = 2500 GeV (left), 3500 GeV (right), $c_{G_{1}}$ = - 0.001  and $g_{Z^{\prime}} \sim 0.6$ for the $U(1)^{\prime}_{L_e + L_\mu + L_\tau}$ model}
\label{fig:channelsall}
\end{figure}

The variation of the relic density with dark matter mass for different values of $f_{b}$ is shown in Fig.\ref{fig:relicLRmutau2500} and Fig.\ref{fig:relicLRmutau3500} for $m_{Z^{\prime}}$ = 2500 GeV and $m_{Z^{\prime}}$ = 3500 GeV respectively, while the mass of the axi-Higgs is related to the symmetry breaking scale $f_b$, as given in Eq.\ref{eq:psuedoGoldstonemass}, where the parameter $c_{G_{1}}$ is taken to be $- 0.001$. 
%We observe that, for each of the benchmark values of $f_b$, and the relic density reaches a minimum near the axi-higgs resonance, i.e., for $m_{G_{1}} \approx 2 m_{DM}$. Hence we see that the axi-higgs can serve as an efficient mediator for the annihilation of DM. \\
%
%
%
%
\\
The relative contributions of different annihilation channels are shown in Fig.\ref{fig:channelsLRmutau}. From this, we see that the $W^+W^-$ channel serves as the most dominant annihilation channel for dark matter particles. In Tab.\ref{table5mutau}, we summarise the relative contribution of annihilation channels for a set of benchmark values of $m_{G_{1}}$ and corresponding dark matter $m_{DM}$ for which the computed relic density lies within $\Omega h^{2}_{critical} = 0.12 \pm 0.001$.
\\
\begin{table}[!h]
\hspace{2cm} 
  \begin{subtable}{.3\linewidth}
       \begin{tabular}{|l|l|l|}
\hline
{\textbf{$m_{G_{1}}$}} & {\textbf{$m_{DM}$}} & {\textbf{Annihilation}}   \\ 
{\textbf (GeV)}& {\textbf (GeV)}& {\textbf{channels}} \\ \hline
330       & 164.87                    & \begin{tabular}[c]{@{}l@{}}  71\% $\chi \overline{\chi} \rightarrow W^{+} W^{-}$\\25\% $\chi \overline{\chi} \rightarrow \gamma Z$\\4\% $\chi \overline{\chi} \rightarrow ZZ$\end{tabular}                      \\ \hline
420      &    209.59                           & \begin{tabular}[c]{@{}l@{}}  72\% $\chi \overline{\chi} \rightarrow W^{+} W^{-}$\\24\% $\chi \overline{\chi} \rightarrow \gamma Z$\\4\% $\chi \overline{\chi} \rightarrow ZZ$  \end{tabular}         \\ \hline
570       & 284.10                                 & \begin{tabular}[c]{@{}l@{}}  73\% $\chi \overline{\chi} \rightarrow W^{+} W^{-}$\\23\% $\chi \overline{\chi} \rightarrow \gamma Z$\\4\% $\chi \overline{\chi} \rightarrow ZZ$ \end{tabular}          \\ \hline
\end{tabular}
%\label{table5mutau}
    \end{subtable}%
\hspace{2cm}   \begin{subtable}{.3\linewidth}
\begin{tabular}{|l|l|l|}
\hline
{\textbf{$m_{G_{1}}$}} & {\textbf{$m_{DM}$}} & {\textbf{Annihilation}}   \\ 
{\textbf (GeV)}& {\textbf (GeV)}& {\textbf{channels}} \\ \hline
420       & 209.97                    & \begin{tabular}[c]{@{}l@{}}  72\% $\chi \overline{\chi} \rightarrow W^{+} W^{-}$\\24\% $\chi \overline{\chi} \rightarrow \gamma Z$\\4\% $\chi \overline{\chi} \rightarrow ZZ$\end{tabular}                      \\ \hline
600      &    299.81                           & \begin{tabular}[c]{@{}l@{}}  73\% $\chi \overline{\chi} \rightarrow W^{+} W^{-}$\\23\% $\chi \overline{\chi} \rightarrow \gamma Z$\\4\% $\chi \overline{\chi} \rightarrow ZZ$  \end{tabular}         \\ \hline
740       & 369.72
& \begin{tabular}[c]{@{}l@{}}  73\% $\chi \overline{\chi} \rightarrow W^{+} W^{-}$\\23\% $\chi \overline{\chi} \rightarrow \gamma Z$\\4\% $\chi \overline{\chi} \rightarrow ZZ$ \end{tabular}          \\ \hline
\end{tabular}
%\label{table6mutau}
    \end{subtable} 
\caption {Benchmark points from Fig.\ref{fig:channels2500LRmutau} and Fig.\ref{fig:channels3500LRmutau} showing values of $m_{G_{1}}$ and $m_{DM}$ for the $U(1)_{L_{\mu}-L_{\tau}}$ model for $m_{Z^{\prime}}=2500$ GeV (left) and 3500 GeV (right) that satisfy the observed DM relic abundance of the universe. The major annihilation channels of the dark matter and their relative contributions are shown in the third column.}
\label{table5mutau} 
\end{table}
%
%
%%%%%%%%%%%%%%%%%%%%%%%%%%%%%%%%
%
%
Similar to the previous calculations, below, we compute the relic density and contributions of the annihilation channels for the $U^{\prime}_{L_e + L_\mu + L_\tau}$ model. Here, the axi-Higgs's couplings to the charged leptons are not suppressed, and hence lower dark matter masses ($m_{DM} \lesssim$ 50 GeV) are allowed and we get the relic density to be within the critical value. This is shown in Fig.\ref{fig:relicLR} and Fig.\ref{fig:omega3500} for $m_{Z^{\prime}}$ = 2500 GeV and $m_{Z^{\prime}}$ = 3500 GeV respectively. And, the different annihilation channels and their relative contributions are shown in Fig.\ref{fig:channels} and Fig.\ref{fig:channels3500}. While the relative contribution of the bosonic channels shows a similar behaviour as the previous case for heavier dark matter, a new tau lepton channel becomes dominant at lower dark matter masses. Hence, the fermionic annihilation channel also becomes important in the $U^{\prime}_{L_e + L_\mu + L_\tau}$ model. Finally, in Tab.\ref{table5}, we summarise the annihilation channels for a set of benchmark values of $m_{G_{1}}$ and corresponding $m_{DM}$ for which the observed relic density is satisfied.
\begin{table}[!h]
\hspace{2cm} 
  \begin{subtable}{.3\linewidth}
       \begin{tabular}{|l|l|l|}
\hline
{\textbf{$m_{G_{1}}$}} & {\textbf{$m_{DM}$}} & {\textbf{annihilation}}   \\ 
{\textbf (GeV)}& {\textbf (GeV)}& {\textbf{channels}} \\ \hline
70     & 33.94                                 & \begin{tabular}[c]{@{}l@{}}99\% $\chi \overline{\chi} \rightarrow \tau \tau$ \end{tabular} \\ \hline
130      & 64.59                                & \begin{tabular}[c]{@{}l@{}}83\% $\chi \overline{\chi} \rightarrow \tau \tau$\\ 17\% $\chi \overline{\chi} \rightarrow \gamma \gamma$\\ \end{tabular}                        \\  \hline
220       & 108.69                                 & \begin{tabular}[c]{@{}l@{}}  68\% $\chi \overline{\chi} \rightarrow W^{+} W^{-}$\\23\% $\chi\overline{\chi} \rightarrow\gamma \gamma$\\  4\% $\chi \overline{\chi} \rightarrow \tau \tau$ \\4\% $\chi \overline{\chi} \rightarrow ZZ$ \end{tabular}          \\ \hline
320       &    156.75                           & \begin{tabular}[c]{@{}l@{}}79\% $\chi \overline{\chi} \rightarrow W^{+} W^{-}$\\13\% $\chi \overline{\chi}\rightarrow\gamma \gamma$ \\ 8\% $\chi \overline{\chi} \rightarrow Z Z$\\ \end{tabular}         \\ \hline
460       & 226.14                    & \begin{tabular}[c]{@{}l@{}}80\% $\chi \overline{\chi}\rightarrow W^{+} W^{-}$\\10\% $\chi\overline{\chi} \rightarrow\gamma \gamma$ \\ 9\% $\chi \overline{\chi} \rightarrow Z Z$\end{tabular}                                                                \\  \hline
\end{tabular}
    \end{subtable}%
\hspace{2cm}   \begin{subtable}{.3\linewidth}
\begin{tabular}{|l|l|l|}
\hline
{\textbf{$m_{G_{1}}$}} & {\textbf{$m_{DM}$}} & {\textbf{annihilation}}   \\ 
{\textbf (GeV)}& {\textbf (GeV)}& {\textbf{channels}} \\ \hline
80     & 39.07                                 & \begin{tabular}[c]{@{}l@{}}99\% $\chi \overline{\chi} \rightarrow \tau \tau$ \end{tabular} \\ \hline
%130      & 65                               & \begin{tabular}[c]{@{}l@{}}94\% $\chi \chi \rightarrow \tau \tau$\\ 6\% $\chi \chi \rightarrow \gamma \gamma$\\ \end{tabular}                        \\  \hline
190        & 94.77                                 & \begin{tabular}[c]{@{}l@{}} 40\% $\chi \overline{\chi} \rightarrow \tau \tau$ \\  35\% $\chi \overline{\chi} \rightarrow W^{+} W^{-}$\\24\% $\chi \overline{\chi} \rightarrow\gamma \gamma$\\ \end{tabular}          \\ \hline
280       & 138.82                                 & \begin{tabular}[c]{@{}l@{}}76\% $\chi\overline{\chi} \rightarrow W^{+} W^{-}$\\15\% $\chi \overline{\chi} \rightarrow\gamma \gamma$ \\ 7\% $\chi \overline{\chi} \rightarrow Z Z$\\2\% $\chi\overline{\chi}\rightarrow \tau \tau$\end{tabular}         \\ \hline
440       & 217.15                     & \begin{tabular}[c]{@{}l@{}}80\% $\chi \overline{\chi} \rightarrow W^{+} W^{-}$\\11\% $\chi \overline{\chi} \rightarrow\gamma \gamma$ \\ 9\% $\chi \overline{\chi} \rightarrow Z Z$\end{tabular}                                                                \\  \hline
\end{tabular}
    \end{subtable} 
\caption {Few benchmark points from Fig.\ref{fig:channels} and Fig.\ref{fig:channels3500} showing values of $m_{G_{1}}$ and $m_{DM}$ for the $U^{\prime}_{L_e + L_\mu + L_\tau}$ model for $M_{Z^{\prime}}=2500$ GeV (left) and 3500 GeV (right) that satisfy the observed DM relic abundance of the universe. The major annihilation channels and their relative contributions are also given.}
\label{table5} 
\end{table}

\section{Indirect Detection}
\label{sec:indirect}
%Indirect detection seeks to identify observable products or signatures resulting from interactions involving dark matter (DM) particles that are already present in the universe. It studies the high-energy cosmic rays or gamma rays that are produced by the astrophysical bodies in various observatories and telescopes. Certain regions of the universe have higher density of dark matter particles such that they can find each other to co-annihilate to SM particles and  produce $\gamma$ or cosmic rays. Hence, we can predict the flux of the incoming photon or cosmic ray signals from these sources, by calculating the velocity averaged cross-section($\braket{\sigma v_{rel}}$) using the current density and distribution of DM, and look for these signals in experiments. The absence of such expected signals can put bounds on the DM parameter space.\\ 

In the previous section, while computing the relic density, we observed that the annihilation channels for the dark matter contain a significant fraction of photon ($\gamma$) production. Such processes, with very high-energy gamma rays, are actively searched for by instruments like Fermi Gamma-ray Large Area Telescope (\textit{Fermi}LAT)~\cite{Atwood_2009, Ackermann_2015}, Cherenkov Telescope Array~\cite{CTAConsortium:2010umy,CherenkovTelescopeArray:2024osy} and High Energy Stereoscopic System (H.E.S.S.)~\cite{Abdallah_2018}.

While \textit{Fermi}LAT is a satellite-based observatory that detects gamma rays in the energy range from $20$ MeV to over $300$ GeV, H.E.S.S. and CTA are ground-based observatories that detect very high-energy gamma-ray fluxes in the range $E_{\gamma}>300$GeV and $200$GeV$<E_{\gamma}<300$TeV respectively. The experiments measure the photon flux received at Earth from galactic centres, given by,
\begin{equation}
\frac{1}{A}\frac{dN_{\gamma}}{dEdt} \propto \frac{\braket{\sigma v_{rel}}_{X\gamma}}{m_{DM}^2}
\end{equation}
where $\frac{dN_{\gamma}}{dEdt}$ denotes the energy spectrum of photons produced in the annihilation process with the velocity-averaged cross-section $\braket{\sigma v_{rel}}_{X\gamma}$. 
%    \frac{1}{A}\frac{dN_{\gamma}}{dEdt}= \frac{\braket{\sigma v_{rel}}}{m_{DM}^2}\Big( \frac{dN_{\gamma}}{dEdt}\Big)_0 J_{ann}
%and the "J-factor" of the source for annihilation is defined as;
%\begin{equation}
%    J_{ann}=\frac{1}{8\pi}\int dr d\Omega\rho(\Vec{r})^2.
%\end{equation}
This velocity-averaged cross-section can be obtained from \cite{Srednicki:1988ce},
\begin{equation}
    \braket{\sigma v_{rel}}_{X\gamma}=\frac{1}{m_{\chi^2}} \Big[\omega(s)_{X\gamma}-\frac{3}{2}(2{\omega(s)}-4m_{\chi}^2 \omega^{\prime}(s)_{X \gamma})\frac{1}{x_f} \Big]\Big|_{s=4m_{\chi}^2}
\end{equation}
where the function $\omega(s)$ is defined as,
\begin{equation}
\omega(s)_{X \gamma}=\frac{1}{32\pi}\sum_{spins}\Big|\mathcal{M}_{\chi \overline{\chi} \to X \gamma}\Big|^2
\end{equation}
where X is either $\gamma$ or Z in our model.

Here, using the NFW~\cite{Navarro_1997} dark matter density profile, we calculate the velocity-averaged cross-section, $\langle \sigma v_{rel} \rangle$, for the production of photons from DM annihilation in both the anomalous $U(1)^{\prime}$ models considered. For the $U(1)^{\prime}_{L_{\mu}-L_{\tau}}$, as described in Sec.\ref{sec:firstmodel}, $g_{a\gamma\gamma}$ = 0, and hence direct annihilation of DM into a pair of photons is not possible. However, we get photons in the final state from the $\chi \overline{\chi} \to \gamma Z$ channel and from the loop level contributions of $\chi \overline{\chi} \to W^{+}W^{-}$ and $\chi \overline{\chi} \to ZZ$. Among these, we consider only the direct channel, as the rest are higher loop suppressed. On the other hand, in the $U(1)^{\prime}_{L_{e}+L_{\mu}+L_{\tau}}$ model, the contribution of the $\chi \overline{\chi}\to \gamma Z$ annihilation channel vanishes and a non-zero $g_{a\gamma\gamma}$ ensures that the contribution of the $\chi \overline{\chi} \to \gamma \gamma$ channel is significant. 

\begin{figure}[!h]
    \begin{subfigure}{0.3\linewidth}
        \includegraphics[width=9cm,height=8cm]{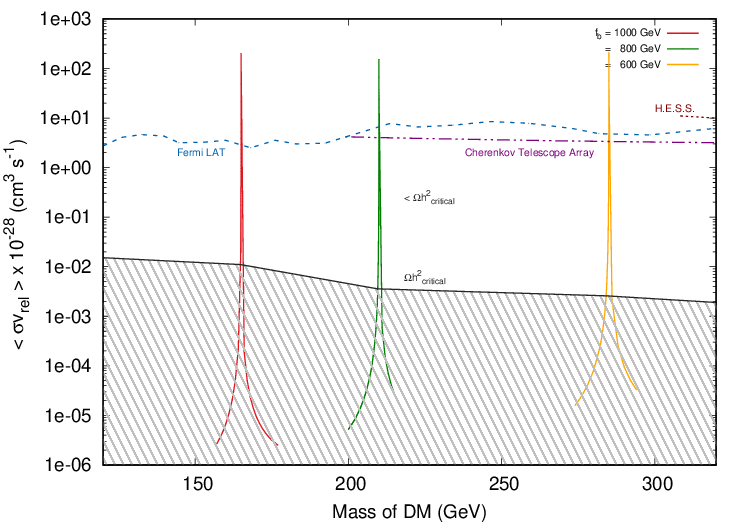}
        \caption{}
		\label{fig:indirect2500mutau}
    \end{subfigure}
    \hspace{4cm}
    \begin{subfigure}{0.3\linewidth}
        \includegraphics[width=9cm,height=8cm]{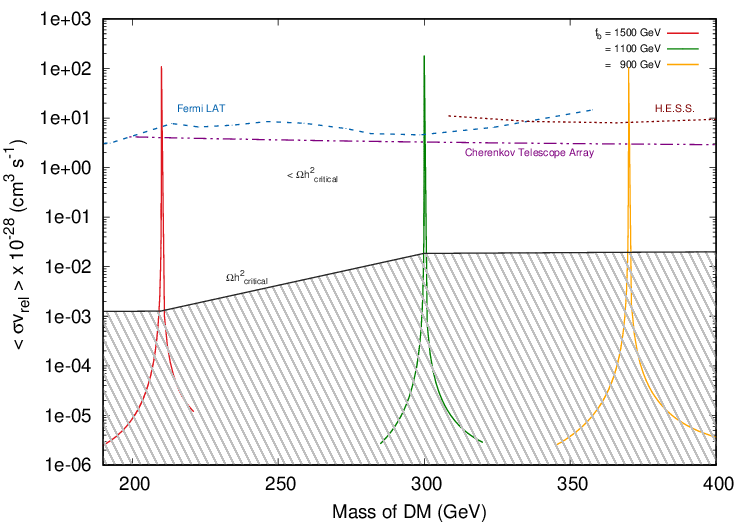}
        \caption{}
	    \label{fig:indirect3500mutau}
    \end{subfigure}
\caption{Velocity-averaged cross-section for $\chi \overline{\chi} \to \gamma Z$ annihilation vs mass of DM plot for the indirect detection of dark matter, for $m_{Z^{\prime}}$ = 2500 GeV (left), 3500 GeV (right), $c_{G_{1}}$ = - 0.001 and $g_{Z^{\prime}} \sim 0.4$ for the $U^{\prime}_{L_\mu - L_\tau}$ model. The grey shaded region is ruled out since the calculated relic density is more than $\Omega h^2_{critical}$ of the universe, while the region above the blue (dashed), purple (dot-dashed) and maroon (dotted) lines are excluded by indirect detection experiments - \textit{Fermi}LAT, CTA and H.E.S.S. - respectively.}
\label{fig:indirectmutau}
\end{figure}

\begin{figure}[!h]
    \begin{subfigure}{0.3\linewidth}
        \includegraphics[width=9cm,height=8cm]{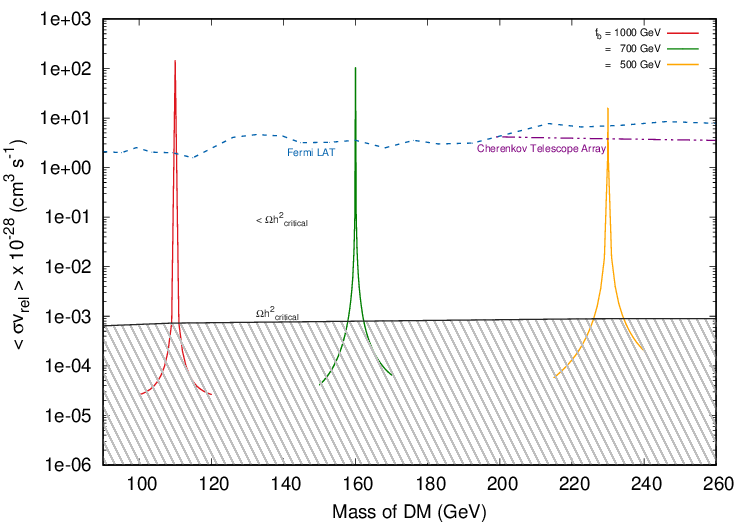}
        \caption{}
		\label{fig:indirect2500all}
    \end{subfigure}
    \hspace{4cm}
    \begin{subfigure}{0.3\linewidth}
        \includegraphics[width=9cm,height=8cm]{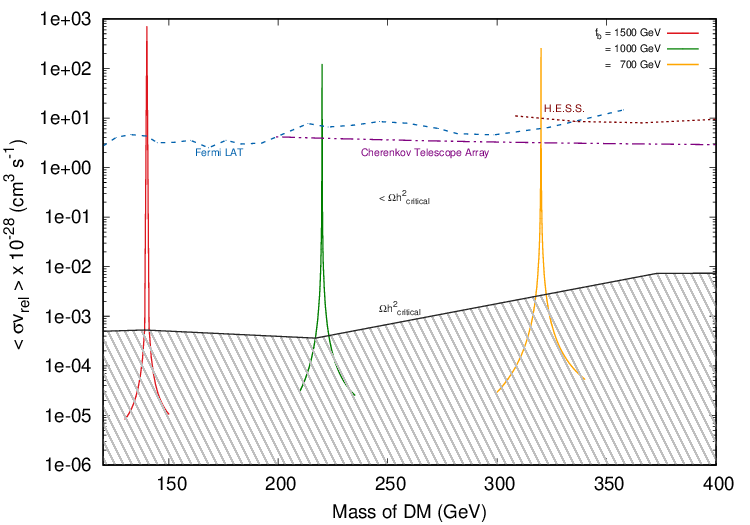}
        \caption{}
	    \label{fig:indirect3500all}
    \end{subfigure}
\caption{Velocity-averaged cross-section for $\chi \overline{\chi} \to \gamma \gamma$ annihilation vs mass of DM plot for the indirect detection of dark matter, for $m_{Z^{\prime}}$ = 2500 GeV (left), 3500 GeV (right), $c_{G_{1}}$ = - 0.001 and $g_{Z^{\prime}} \sim 0.6$ for the $U^{\prime}_{L_e + L_\mu + L_\tau}$ model. The grey shaded region is ruled out since the calculated relic density is more than $\Omega h^2_{critical}$ of the universe, while the region above the blue (dashed), purple (dot-dashed) and maroon (dotted) lines are excluded by indirect detection experiments - \textit{Fermi}LAT, CTA and H.E.S.S. - respectively.}
\label{fig:indirectall}
\end{figure}

For lower mass dark matter in the range of 10 GeV to 200 GeV, the \textit{Fermi}LAT\cite{Ackermann_2015} data provides the strongest upper bounds, while for higher masses ($m_{DM} >$ 200 GeV), the strongest bounds come from CTA\cite{CherenkovTelescopeArray:2024osy}. The bounds from the H.E.S.S.\cite{Abdallah_2018} data are also considered. Upon comparing the $\langle \sigma v_{rel} \rangle_{X\gamma}$ calculated for our models with the experiments, we see that the parameter space that satisfies the observed relic density is allowed by these experiments. The results are shown in Fig.\ref{fig:indirectmutau} for the anomalous $U(1)^{\prime}_{L_{\mu}-L_{\tau}}$ and in Fig.\ref{fig:indirectall} for the anomalous $U(1)^{\prime}_{L_{e}+L_{\mu}+L_{\tau}}$ model. The grey shaded area corresponds to the region where the relic density is greater than the observed critical value, and hence ruled out. And, the region above the dashed, dash-dotted, and dotted lines are ruled out by \textit{Fermi}LAT, CTA, and H.E.S.S. data respectively. 

\section{Conclusion}
\label{sec:conclusion}
One of the simplest beyond Standard Model scenarios is an extension of the Standard Model gauge group with an $U(1)^{\prime}$. Such models naturally exist and are anomalous in orientifolds~\cite{Ibanez:1998qp}. The gauge anomalies present in these extensions are usually cancelled, in string theory, by the Green-Schwarz mechanism. In low energy effective quantum field theory framework, the Green-Schwarz term arises from the well-known Wess-Zumino interaction, $F_I\wedge F_J$, with a St{\"u}ckelberg axion. This Wess-Zumino effective action provides a counterterm that cancels the anomalies rendering the model gauge invariant. Indeed, rearranging and integrating out the axions in the Wess-Zumino case generates the Green-Schwarz effective action, they are non-local and the two mechanisms are different albeit having the same origin.

The aforementioned St{\"u}ckelberg axion readily mixes with other pseudo-Goldstone bosons in the theory to form axi-Higgs, which are massive and remain in the spectra after gauge fixing. Here, in this article, we are concerned with the possibility that these axi-Higgs could mediate dark matter annihilation to Standard Model fields. To this end, we consider an anomalous $U(1)^{\prime}$ extension to the Standard Model along with a complex scalar field and fermionic dark matter. The complex scalar field is necessary to give the dark matter its mass. This simple extension is strikingly different from other models found in literature, owing to the fact that we do not have a multitude of fields whose charges under $U(1)^{\prime}$ are carefully distributed. 

For concreteness, we considered two models, an anomalous $U(1)^{\prime}_{e_\mu L_\mu -e_\tau L_\tau}$ and anomalous $U(1)^{\prime}_{\gamma_5(L_e+L_\mu+L_\tau)}$. Quarks are considered to be singlets under the abelian gauge groups, to make the work simple and clear without other constraints from hadron physics. Nevertheless, the framework can be extended to other scenarios with quark couplings as well. In these models, we compute the relic density arising from the annihilation channels $\chi\overline{\chi} \to W^+W^-,\gamma\gamma, Z \gamma, ZZ$, and $\tau\tau$. These results are shown in Fig.\ref{fig:relicLRmutau} and Fig.\ref{fig:omegaall}, and the relative contributions of these channels are computed as a function of the mediator mass and are shown Fig.\ref{fig:channelsLRmutau} and Fig.\ref{fig:channelsall}. A few benchmark points are given in Tab.\ref{table5mutau} and Tab.\ref{table5}.

Since quarks are not charged under $U(1)^{\prime}$, direct detection experiments do not constrain the models at tree-level. On the other hand, since the $\gamma$ production channels exist, constraints from {\it Fermi}LAT, Cherenkov Telescope Array and H.E.S.S. become important and the allowed parameter space is shown in Fig.\ref{fig:indirectmutau} and Fig.\ref{fig:indirectall}. Thus, we show here that minimal extensions of the Standard Model, which are anomalous, can host interesting dark matter phenomenology.

\section*{Acknowledgments}
M.T.A. acknowledges the financial support of DST through the INSPIRE Faculty grant 
 DST/INSPIRE/04/2019/002507. 

\bibliographystyle{unsrt} 
\bibliography{jcap} 

\end{document}